\documentclass[letterpaper,journal]{IEEEtran}

\usepackage{amsmath,amsfonts,amssymb,amsthm,mathtools,bm}
\usepackage{cite}

\usepackage{color}
\usepackage{xcolor}

\usepackage{graphicx}
\usepackage{epstopdf}
\usepackage{float}
\usepackage{stfloats}

\usepackage[caption=false,font=normalsize,labelfont=sf,textfont=sf]{subfig}
\usepackage[justification=centering]{caption}

\usepackage{booktabs}
\usepackage{multirow}
\usepackage{tabularx}
\usepackage{threeparttable}
\usepackage{verbatim}
\usepackage{setspace}
\usepackage[most]{tcolorbox}

\usepackage[linesnumbered,ruled,vlined]{algorithm2e}
\SetKwInput{KwInput}{Input}
\SetKwInput{KwOutput}{Output}
\SetKwBlock{Where}{where}{}
\SetKwComment{tcp}{// }{}
\DontPrintSemicolon  

\usepackage[hyphens]{url} 
\usepackage[ breaklinks=true]{hyperref} 

\allowdisplaybreaks[0] 

\usepackage{fancyhdr}

\def\BibTeX{{\rm B\kern-.05em{\sc i\kern-.025em b}\kern-.08em
    T\kern-.1667em\lower.7ex\hbox{E}\kern-.125emX}}

\usepackage{balance}
\begin{document}

\title{Velocity-Adaptive Access Scheme for Semantic-Aware Vehicular Networks: Joint Fairness and AoI Optimization }

\author{Xiao Xu, Qiong Wu,~\IEEEmembership{Senior Member,~IEEE}, Pingyi Fan,~\IEEEmembership{Senior Member,~IEEE}, \\
Kezhi Wang,~\IEEEmembership{Senior Member,~IEEE}, Nan Cheng,~\IEEEmembership{Senior Member,~IEEE},\\ Wen Chen,~\IEEEmembership{Senior Member,~IEEE}, 
and Khaled B. Letaief,~\IEEEmembership{Fellow,~IEEE}

\thanks{This work was supported in part by Jiangxi Province Science and Technology Development Programme under Grant No. 20242BCC32016, in part by the National Natural Science Foundation of China under Grant No. 61701197 and 62531015, in part by the Basic Research Program of Jiangsu under Grant BK20252084, in part by the National Key Research and Development Program of China under Grant No. 2021YFA1000500(4), in part by the Shanghai Kewei under Grant 24DP1500500, in part by the Research Grants Council under the Areas of Excellence Scheme under Grant AoE/E-601/22-R and in part by the 111 Project under Grant No. B23008. 

Xiao Xu and  Qiong Wu  are with the School of Internet of Things Engineering, Jiangnan University, Wuxi 214122, China, and also with the School of Information Engineering, Jiangxi Provincial Key Laboratory of Advanced Signal Processing and Intelligent Communications, Nanchang University, Nanchang 330031, China (e-mail:  xuxiao@stu.jiangnan.edu.cn, qiongwu@jiangnan.edu.cn). (Corresponding author: Qiong Wu.)


Pingyi Fan is with the Department of Electronic Engineering, State Key laboratory of Space Network and Communications,Beijing National Research Center for Information Science and Technology, Tsinghua University, Beijing 100084, China (email: fpy@tsinghua.edu.cn).

Kezhi Wang is with the Department of Computer Science, Brunel University, London, Middlesex UB8 3PH, U.K. (email: Kezhi.Wang@brunel.ac.uk)

Nan Cheng is with the State Key Laboratory of ISN and the School of Telecommunications Engineering, Xidian University, Xi’an 710071, China (e-mail: dr.nan.cheng@ieee.org).

Wen Chen is with the Department of Electronic Engineering, Shanghai Jiao Tong University, Shanghai 200240, China (e-mail: wenchen@sjtu.edu.cn).

Khaled B. Letaief is with the Department of Electrical and Computer Engineering, the Hong Kong University of Science and Technology, Hong Kong (e-mail: eekhaled@ust.hk).

}
}


\maketitle

\begin{abstract}
In this paper, we address the problem of fair access and Age of Information (AoI) optimization in 5G New Radio (NR) Vehicle to Everything (V2X) Mode 2. Specifically, vehicles need to exchange information with the road side unit (RSU). However, due to the varying vehicle speeds leading to different communication durations, the amount of data exchanged between different vehicles and the RSU may vary. This may poses significant safety risks in high-speed environments.
To address this, we define a fairness index  through tuning the selection window of different vehicles and consider the image semantic communication system to reduce latency. However, adjusting the selection window may affect the communication time, thereby impacting the AoI. Moreover, considering the re-evaluation mechanism in 5G NR, which helps reduce resource collisions, it may lead to an increase in AoI. We analyze the AoI using Stochastic Hybrid System (SHS) and construct a multi-objective optimization problem to achieve fair access and AoI optimization. Sequential Convex Approximation (SCA) is employed to transform the non-convex problem into a convex one, and solve it using convex optimization. We also provide a large language model (LLM) based algorithm. The scheme’s effectiveness is validated through numerical simulations.

\end{abstract}


\IEEEpeerreviewmaketitle

\section{Introduction}
\label{sec1}
\IEEEPARstart{W}{ith} the advancement of vehicular networks, the update speed of vehicular applications has significantly accelerated \cite{intro1,arx1,arx2}. Intelligent driving technologies is one of the most important technologies \cite{intro2,arx3,arx4}. However, the massive computational tasks generated by these applications and technologies pose great challenges to vehicles with limited computing resources \cite{intro3}. Moreover, in high-speed environments, the lack of timely information and communication delays can be dangerous. To address this, vehicles use Mobile Edge Computing (MEC) to communicate with Road Side Units (RSUs) in order to get exact information in a timely manner \cite{intro4}.

Vehicular applications are no longer limited to text data transmission. Instead, they often involve image data \cite{intro5,aa1}. Intelligent driving technologies require vehicles to collect precise environmental information via onboard sensors, such as high-definition cameras \cite{intro6,arx11}. Transmitting large volumes of image data to the RSU can be time-consuming, and semantic communication has emerged as one potential solution \cite{intro7}.

The goal of communication is to allow the receiver to understand the sender’s intent \cite{intro8,arx5}. Semantic communication achieves this by extracting semantic information from the sender’s data and transmitting only that. In image semantic systems, semantic information is extracted from images and transmitted as textual information, which avoids transmitting the entire image and focuses on transmitting the core content. Therefore, introducing semantic communication systems into vehicular networks can greatly reduce the difficulty of data transmission and the Age of Information (AoI).

Currently, vehicles use the 5G New Radio (NR) Vehicle to Everything (V2X) protocol introduced in 3GPP Release 16 for data transmission \cite{5G}. This protocol has two modes: Mode 1 and 2. Compared to Mode 1, Mode 2 does not require full network coverage, making it more flexible and better suited for vehicle to RSU communication. Hence, vehicles typically use the scheduling-based Semi-Persistent Scheduling (SPS) mechanism in Mode 2 for resource allocation \cite{intro9,arx6}.

However, in real-world vehicular networks, RSU coverage is limited, and vehicle speeds often differ across lanes. As a result, vehicles at different speeds spend different amounts of time within the RSU’s communication range, leading to unequal opportunities for data exchange. Faster vehicles may receive less useful information than slower ones, which can pose risks when all vehicles are operating in the same region.

Additionally, Mode 2 introduces a re-evaluation mechanism which allows vehicles to reassess communication resources after initial selection to avoid collisions \cite{intro10,arx7,arx8}. While this mechanism reduces resource conflicts, the change in transmission time may increase the AoI-critical metric in high-speed vehicular environments \cite{intro11,arx9,arx10}. High AoI can cause delays or outdated decisions in vehicular networks. Thus, it is necessary to consider the mechanism's influence on AoI.

In this work, we define a fairness index for semantic image communication systems and dynamically adjust each vehicle’s selection window size according to its velocity to realize fairness access. Meanwhile, considering how selection window adjustments and the re-evaluation mechanism influence AoI, we model AoI using a Stochastic Hybrid System (SHS). We construct a multi-objective optimization task, then solve it using Sequential Convex Approximation (SCA) method and Large Language Model (LLM) based multi-Objective Evolutionary Algorithm based on Decomposition (MOEA/D). As far as we know, no existing work has comprehensively considered both fair access in semantic image systems and  the re-evaluation mechanism's influence on AoI, which serves as the motivation for this work.

This paper makes the following key contributions\footnote{Source code can be found at : \url{https://github.com/qiongwu86/Velocity-Adaptive-Access-Scheme-for-Semantic-Aware-Vehicular-Networks-Joint-Fairness-and-AoI}}:
\begin{itemize}
\item[1)] We consider the fair access problem caused by speed differences in a semantic image communication system. A fairness index is defined to measure access fairness, and selection window sizes are tuned according to vehicles' velocity to achieve fairness access.
\item[2)]  We analyze how changes in the selection window affect AoI, while also considering the unique re-evaluation mechanism of 5G NR. Using an SHS model, we characterize the relationship between AoI and the selection window size under re-evaluation.
\item[3)] We design a multi-objective optimization task to optimize fair access and AoI by dynamically adjusting selection windows based on speed. We provide two algorithms including SCA and LLM based MOEA/D. Numerical simulations confirm the validity of our method.
\end{itemize}

The paper is organized as follows:
Section \ref{sec2} covers related work. Section \ref{sec3} outlines the system model, while Section \ref{sec4} presents the definition of the fairness index. Section \ref{sec5} models the AoI. Section \ref{sec6} constructs a multi-objective optimization problem and slove it. Section \ref{sec7} analyzes the numerical simulation results. Section \ref{sec8} concludes the paper.

\section{Related Works}
\label{sec2}
\subsection{Fairness access}
We first review the previous work related to fair access.
In \cite{fair1}, Trung \textit{et al.} achieved fairness among all users in a distributed joint sensing and communication (JSC) system by optimizing the bandwidth allocation between the sensing and communication functions of the JSC nodes.
In \cite{fair2}, Fidan \textit{et al.} analytically addressed the max-min fairness problem in Software-Defined Radio Access Networks (SD-RANs) and proposed solutions for two conditions: one aiming to provide max-min fairness for all network users , and the other targeting max-min fairness in terms of throughput both base stations and  users.
In \cite{fair3}, Wang \textit{et al.} investigated multi-user multiple-input multiple-output (MIMO) technology for licensed spectrum sharing, proposing a fair spectrum allocation scheme between two mobile network operators in a multi-cell, multi-user MIMO network. Fairness is maintained by allocating spectrum to each operator in proportion to its contribution.
None of the above works have considered the access fairness issue in vehicular networks.

In vehicular networks, there are also some related studies.
In \cite{wan}, Wan \textit{et al.} considered the fair access and AoI optimization for vehicle in the 802.11p simultaneously, and designed a scheme that dynamically tunes the contention window (CW) based on speed in order to achieve fair access and optimize AoI, and finally adopted a heuristic algorithm to solve the problem.
In \cite{fair4}, Wu \textit{et al.} also addressed the issues of fair access and AoI for vehicle nodes under the IEEE 802.11p protocol, and employed an extended Deep Q-learning (DQN) method to allocate optimal CW values for the nodes.
However, these studies only focus on the IEEE 802.11p protocol. Research related to 5G NR V2X remains lacking, which motivates our work.

\subsection{Age of information}  

We continue to review related work on AoI under 5G NR.
In \cite{reaoi1}, Rolich \textit{et al.} pointed out that while SPS improves resource utilization efficiency, it does not consistently enhance AoI performance. They revealed the fundamental trade-off between reliability (PDR) and timeliness (AoI), emphasizing the need for careful persistence management in vehicular communication systems.
In \cite{reaoi2}, Alexey \textit{et al.} developed an SPS parameter adaptation method aimed at reducing the average AoI of cooperative awareness messages. Built upon analytical modeling of SPS, the method’s performance is validated via simulation.
In \cite{reaoi3}, Ekaterina \textit{et al.} analyzed the influence of 5G NR system configurations on application-level AoI performance. To achieve this, they used queuing theory and stochastic geometry tools to quantify the mean and distribution of PAoI and sojourn time, taking into account traffic arrival patterns and the coefficient of variation and autocorrelation in service characteristics within the 5G NR system.
In \cite{reaoi4}, Wang \textit{et al.} noted that conventional broadcast-based data dissemination often leads to congestion, as it requires participation from a large number of vehicles, thereby resulting in a higher AoI. To address this, they leveraged mobile edge computing and proposed a new unicast-based dissemination method to reduce AoI.
In \cite{reaoi5}, Song \textit{et al.} discussed the joint optimization of AoI and energy consumption in 5G NR and employed Deep Reinforcement Learning (DRL) to address the problem. However, they did not provide an discussion on AoI modeling or the impact of the re-evaluation mechanism.
In conclusion, the above works did not consider the impact of selection window size or the re-evaluation mechanism on AoI. They also overlooked the issue of fair access in vehicular networks and did not model AoI using SHS.

\section{System Model}
\label{sec3}
In this section, we introduces the system model which includes the scenario model, the semantic communication model and the SPS in 5G NR.
\begin{figure*}[htbp]
	\centering
	\includegraphics[width=\linewidth, scale=1.00]{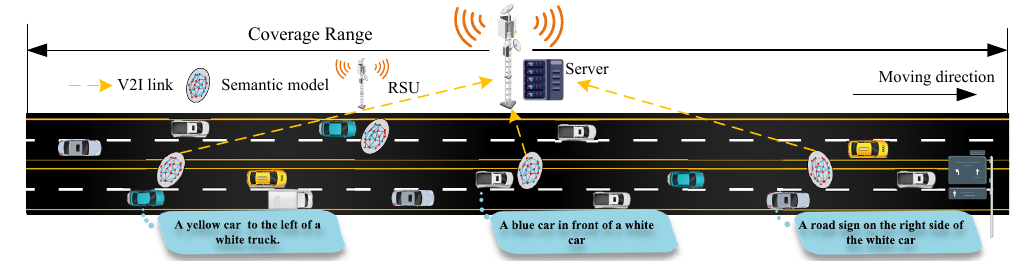}
	\caption{Scenario Model}
	\label{fig1}
	\vspace{-0.7cm}
\end{figure*}

\subsection{Scenario Model}
As shown in Fig. \ref{fig1}, we focus on a  multi-lane highway scenario comprising $N$ lanes. RSUs equipped with computationally capable servers, are positioned along the roadside. The number of vehicles covered by an RSU is set as $N_v$. Upon entering the communication range of an RSU, vehicles attempt to establish communication with it. This paper specifically focuses on the scenario where vehicles transmit images to the RSU. Typically, vehicles exchange data with the RSU via the 5G NR Vehicle to Infrastructure (V2I) Mode 2 protocol. Given the large data volumes associated with image transmission, semantic communication is employed to reduce the amount of transmitted data. We assume that vehicles within the same lane maintain a constant speed and constant inter-vehicle spacing, while vehicles in adjacent lanes exhibit a speed difference of at least 3.6 m/s. Similar to \cite{wan}, this paper considers only the uplink transmission, as the useful data is minimal, the downlink capacity is substantially greater than the uplink. The initial distribution of vehicles follows a Poisson distribution.
\subsection{Semantic Model}
Following the model in \cite{semantic}, we categorize semantic information into nodes and the relationships between nodes. Each node represents an object, such as a vehicle, pedestrian, etc. The image's semantic data can be decomposed into multiple semantic triples. A semantic triple can be represented as: [Vehicle 1] [in front of] [Vehicle 2]. Here, vehicle 1 and vehicle 2 are two nodes, and is in front of denotes the relationship between them. Consequently, multiple semantic triples can be used to describe a single image. We employ the scene graph extraction model proposed in \cite{semantic2} to obtain semantic information from the raw images captured by vehicles. This extracted semantic information is then transmitted in the form of triples. The semantic triples corresponding to image $I_k$ transmitted by vehicle $v$ can be represented as:
\begin{equation}
	S_{vk}^{{}}=\left\{ s_{vk}^{1},s_{vk}^{2},\cdots ,s_{vk}^{n},\cdots ,s_{vk}^{{{N}_{vk}}} \right\},
	\label{eq001}
\end{equation}
where $s_{vk}^{n}=\left( o_{vk,i}^{n},r_{vk,ij}^{n},o_{vk,j}^{n} \right)$, where \(o_i\) and \(o_j\) denote objects \(i\) and \(j\) in image \(k\), respectively, and \(r_{ij}\) describes the relationship between objects \(i\) and \(j\). We remark that \(r_{ij} \neq r_{ji}\) due to the directional nature of relationships. \(N_{vk}\) denotes the quantity of semantic triples in image \(k\). Thus, a semantic triple \(s\) can be formally defined as \(s = (o_i, r_{ij}, o_j)\), and the collective set \(S = \{s_1, s_2, ..., s_{N_{vk}}\}\) constitutes the complete semantic representation of an image.  
\begin{figure}[bp]
	\centering
	\includegraphics[width=\linewidth, scale=1.00]{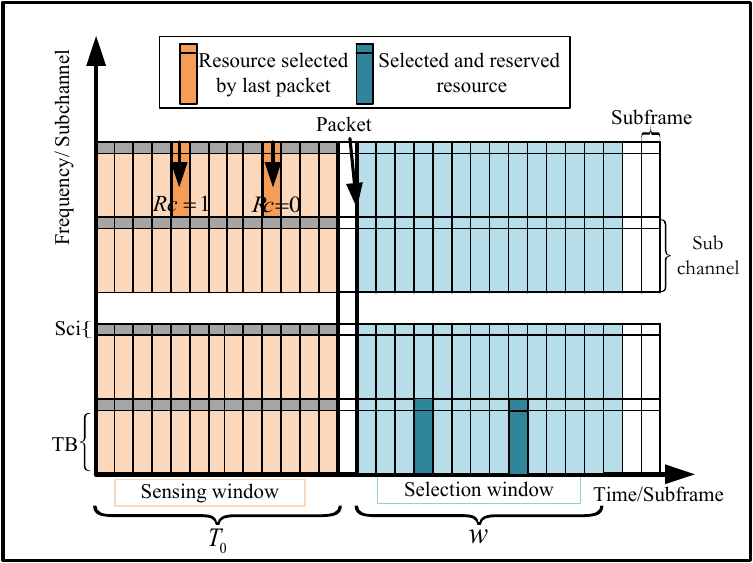}
	\caption{SPS Model}
	\label{fig2}
	\vspace{-0.7cm}
\end{figure}
  
To quantify the data volume of semantic information,  \(L(x)\) was defined as the character count in a triple component \(x\). Consequently, the  semantic data volume for image \(I_k\) transmitted by vehicle \(v\) is given by:  
\begin{equation}
	L(S_{vk}^{{}})=\sum\limits_{n=1}^{{{N}_{vk}}}{(Z(o_{vk,i}^{n})}+Z(r_{vk,ij}^{n})+Z(o_{vk,j}^{n})).
	\label{eq002}
\end{equation}

Semantic communication typically employs semantic similarity as the performance metric. Therefore, it is necessary to assess the similarity between the semantic data and the original image. Following \cite{semantic}, we adopt an image-to-graph semantic similarity metric, defined as:
\begin{equation}
	\xi (S_{vk}^{{}},{{G}_{vk}})=\frac{\left\| \sum\limits_{n=1}^{{{N}_{vk}}}{\overline{D(s_{vk}^{n})}D{{({{G}_{vk}})}^{T}}\overline{D(s_{vk}^{n})}} \right\|}{\left\| D({{G}_{vk}}) \right\|},
	\label{eq003}
\end{equation}
where \(D(\cdot)\) denotes the vectorization function of a pre-trained Deep Neural Network (DNN), then orthogonalize it as $\overline{D(\cdot)}$ by the GramSchmidt algorithm.  Through this pre-trained DNN, we obtain the vectorized semantic data and the image.

\subsection{SPS Model}
In accordance with the V2X protocol specified in 3GPP Release 16, vehicles typically perform resource allocation and communicate with other devices. Similar to Mode 4 in LTE-V2X, vehicles operating in Mode 2 of 5G NR can autonomously allocate resources from the communication resource pool without network assistance. Under Mode 2, vehicles typically employ a SPS scheme for resource scheduling. As illustrated in Fig. \ref{fig2}, communication resources are divided into multiple subchannels in the frequency domain. Each subchannel consists of a Scheduling Control Information (SCI) field and a transport channel, where SCI carries control information and the transport channel transmits Transport Blocks (TBs). A vehicle reserves a subchannel based on its Resource Reservation Interval (RRI) for transmitting its next data packet. When selecting a new resource, the vehicle enters an exclusion phase within a selection window, utilizing resource information obtained during a sensing window. During this exclusion phase, the vehicle must exclude resources already reserved by other vehicles or those whose SCI-indicated Reference Signal Received Power (RSRP) exceeds a predefined threshold. After exclusion, the vehicle chooses a resource at random from the remaining available resource list for communication.  
  
Notably, unlike LTE V2X, NR V2X introduces a novel mechanism — the re-evaluation mechanism. Upon adopting this mechanism, a vehicle continues sensing resource information during the original selection window. If the vehicle detects that its selected resource becomes unavailable, it defines a new selection window, performs the resource exclusion procedure again, and subsequently selects a new resource.
\section{Fairness Index}
\label{sec4}

This section considers the transmission rate, semantic similarity, and transmission collision probability under 5G NR in an image semantic communication system, and establishes a fairness index to assess the extent to which vehicles achieve fair access. The important parameters employed in this paper are listed in Table \ref{tab1}.

\begin{table}\footnotesize
	
	\caption{Important Symbols }
	\label{tab1}
	\centering
	\begin{tabular}{|c|p{6.6cm}|}
		\hline
		\textbf{Symbol} &\textbf{Description}\\
		\hline
		$V_v$  & Vehicle $v$'s speed.\\
		\hline
		$N_v$  &  The number of vehicles.\\
		\hline
		$R$  &  RSU's communication range. \\
		\hline
		$G_{max}$ & Maximum iteration number. \\
		\hline
		$RRI$ & Source selection interval.\\
		\hline
		$B$ & Bandwidth. \\
		\hline
		$\mu$ & Subcarrier spacing factor. \\
	    \hline
		$V_{min}$   & The lower bound of vehicle's speed.  \\
		\hline
		$V_{max}$   & The upper bound of vehicle's speed.   \\
		\hline
		$I_v$ & The expected data volume transmitted by vehicle $v$. \\
		\hline
		$P_v$ &The number of images transmitted by vehicle $v$ \\
		\hline	
		$D_p$ & \multicolumn{1}{m{6.6cm}|}{The average successfully transmitted data size per image.} \\
		\hline
		$T_v$ &  Vehicle $v$'s dwell time within the RSU’s coverage area. \\
		\hline
		$T_v^p$ &The average transmission time per image for vehicle $v$. \\
		\hline
		\( p_v \) &   Vehicle $v$'s power.  \\
		\hline
		\( PR_v \)  &\multicolumn{1}{m{6.6cm}|}{The successful communication probability between vehicle $v$ and the RSU.} \\
		\hline
		\( \delta_{v,j} \) & The transmission collision probability. \\
		\hline
		\( B_l \)  & The number of bits per transmitted character. \\
		\hline
		\( ISS_v \) & The average semantic similarity for vehicle $v$. \\
		\hline
        \(h_v \) & The channel gain between vehicle $v$ and the RSU.  \\
		\hline
		\( G_{fair}^v \) & The fairness index of vehicle $v$. \\
		\hline
		$G_{fair}$ & The averaged network’s fairness.  \\
		\hline
		
		\( \boldsymbol{A(t)} \)  & The set of current AoI.\\
		\hline
		\( S(t) \) &The current link state. \\
		\hline
		
		$A_r(t)$ & The AoI of the data stored at the RSU. \\
		\hline
		\( A_v(t) \)  &The AoI at link $v$. \\
		\hline
		\( \boldsymbol{{{\psi }_{{{l}}}}} \) & The transition mapping matrix for $l$. \\
		\hline
		\( \eta_l  \)  &The transition rate of $l$. \\
		\hline
		\( R_v \) & The average transmission failure rate. \\
		\hline
		\( H_v \) &The average service rate. \\
		\hline
		\( E_v \) & The average re-evaluation rate.\\
		\hline
		\( \boldsymbol{{\bar{Q}}}_{sl} \) &  The steady state correlation. \\
		\hline
		\( \bar{\Delta}_k  \) &The AoI of link $k$ . \\
		\hline
		$\boldsymbol{b_s}$ &  A binary differential vector coefficient. \\
		\hline
		${{\bar{\pi }}_{s}}$ &  The steady-state probability of state $s$. \\
		\hline
	\end{tabular}
\end{table}
\subsection{Transmission rate}

Fair access implies that the data volume transmitted by each vehicle should be comparable. Since vehicles transmit numerous images during operation, we have:
\begin{equation}
	 E({{I}_{v}})=C,
	\label{eq1}
\end{equation}
where \(I_v\) denotes the expected data volume transmitted by vehicle \(v\). Due to probabilistic transmission failures and slight variations in the semantic information size across images, we consider the expected value. \(C\) is a constant.  
Furthermore, \(I_v\) can be expressed as:
\begin{equation}
	{{I}_{v}}={{P}_{v}}\cdot {{D}_{p}},
	\label{eq2}
\end{equation}
where \(P_v\) refers to the total images sent by vehicle \(v\). \(D_p\) denotes the average successfully transmitted data volume per image. \(P_v\) can be further expressed as:
\begin{equation}
{{P}_{v}}=\frac{{{T}_{v}}}{T_{v}^{p}},
\label{eq3}
\end{equation}
where \(T_v\) denotes the period vehicle $v$ remains under the RSU's coverage. \(T_{v}^P\) represents the average transmission time per image for vehicle \(v\). \(T_v\) is given by:
\begin{equation}
{{T}_{v}}=\frac{R}{{{V}_{v}}},
\label{eq4}
\end{equation}
where \(R\) denotes RSU's coverage. \(V_v\) denotes vehicle \(v\)'s speed.  \(T_{v}^P\) can be expressed as:
\begin{equation}
T_{v}^{p}=\frac{{I}}{B\cdot {{\log }_{2}}(1+\frac{{{p}_{v}}\cdot {{h}_{v}}\cdot {{d}_{v}}^{-\partial }}{{{\sigma }^{2}}})\cdot P{{R}_{v}}},
\label{eq5}
\end{equation}
where \(p_v\) represents vehicle $v$'s power, \(h_v\) indicates the channel gain between vehicle \(v\) and the RSU, \(d_v\) denotes their separation distance, \(\alpha\) denotes the path loss exponent, \(\sigma^2\) signifies the noise power, \(PR_v\) represents the successful communication probability between vehicle \(v\) and the RSU, and \(I\) denotes the number of bits per image, \(B\) denotes the bandwidth. According to \cite{armodel}, the channel model adopts an autoregressive model, i.e.:
\begin{equation}
{{h}_{v}}={{\rho }_{v}}{{h}_{v}}^{\prime }+e\sqrt{1-\rho _{v}^{2}},
\label{eq8}	
\end{equation}
where ${\rho }_{v}$ is the autocorrelation coefficient, ${{h}_{v}}^{\prime }$ denotes the channel gains from the preceding slot, with $e$  following a Gaussian distribution. 

\subsection{Successful receiving probability}
We now discuss \(PR_v\). Assuming all vehicles support full-duplex communication, \(PR_v\) can be expressed as:  
\begin{equation}
	P{{R}_{v}}=\prod\limits_{j\ne v}{(1-{{\delta }_{v,j}})},
	\label{eq10}	
\end{equation}
where ${{\delta }_{v,j}}$ denotes the transmission collision probability for vehicle \(v\) and vehicle \(j\).
which denotes that the vehicles experience contention over Physical Resource Blocks (PRBs). Such contention occurs when several transmitters, operating within overlapping time frames, attempt to occupy identical PRBs. Based on \cite{prr1}, ${{\delta }_{v,j}}$ is expressed as:
\begin{equation}
	{{\delta }_{v,j}}={{P}_{O}}\cdot{{P}_{SH|O}}\cdot\frac{{{C}_{Ca}}}{N_{Ca}^{2}},
	\label{eq11}	
\end{equation}
where $P_O$ quantifies the chance of temporal overlap between the selection windows of distinct vehicles. $P_{SH|O}$ denotes the conditional probability that they select resources from this overlapping region. When an overlap happens, $C_{Ca}$ refers to the number of shared candidate PRBs, and $N_{Ca}$ indicates the  mean count of PRBs within the candidate pool. The associated conditional probability $P_O$ is defined as:
\begin{equation}
{{P}_{O}}=\frac{{{w}_{v}}+{{w}_{j}}+1}{1000\cdot {{2}^{\mu }}\cdot RRI},
	\label{eq12}	
\end{equation}
where $w_v$ and $w_j$ indicate the selection window sizes of vehicles $v$ and $j$, respectively. $RRI$ denotes the interval between consecutive resource selections, and $\mu$ is the subcarrier spacing factor.
${P}_{SH|O}$ is given by:
\begin{equation}
	{{P}_{SH|O}}={{(\frac{{{N}_{Sc}}\cdot{{N}_{Sh}}}{{{N}_{r}}})}^{2}}.
	\label{eq13}	
\end{equation}
$N_{Sc}$ represents the total subchannels, and $N_{Sh}$ refers to the resources shared across the overlapped window. $N_r$ denotes the overall resource count. $N_{Sh}$ is defined as:
\begin{equation}
	{{N}_{Sh}}=\frac{({{w}_{v}}+1)({{w}_{j}}+1)}{{{w}_{v}}+{{w}_{j}}+1}.
	\label{eq14}	
\end{equation}

\subsection{Fair index}
From Eq. (\ref{eq002}), we obtain the semantic information volume per single image. However, vehicles transmit numerous images within the RSU’s communication range, which can incur significant latency in high-mobility vehicular environments. To mitigate computational overhead, we precompute the average semantic information volume per image. Consequently, \(I\) can be further expressed as:  
\begin{equation}
{I}=\frac{{{B}_{l}}\cdot \sum\limits_{k=1}^{K}{L({{S}_{vk}})}}{K},
\end{equation} 
where \(K\) denotes the part of images in Visual Genome (VG) dataset \cite{dataset}, and \(B_l\) represents the number of bits per transmitted character.  
 
Through Eq. (\ref{eq3}), we obtain the number of images vehicle \(v\) can transmit within the RSU’s communication range. However, since semantic communication is employed, we must ensure the extracted semantic information accurately represents the original images. To evaluate semantic communication quality within the RSU coverage, we define the average semantic similarity for vehicle \(v\) as:
\begin{equation}
IS{{S}_{v}}=\frac{\sum\limits_{k=1}^{k=G}{\xi (S_{vk}^{{}},{{G}_{vk}})}}{G}.
\label{eq17}	
\end{equation}
Considering that calculating the similarity of all images may consume extensive computational resources, we set $G$ as the upper limit on the number of similarity calculations.
Therefore, the average successfully transmitted data size per image is:
\begin{equation}
{{D}_{p}}=IS{{S}_{v}}\cdot {I}.
\label{eq18}	
\end{equation}
Accordingly, Eq. (\ref{eq1}) can be restated in the following form:
\begin{equation}
\frac{R\cdot \sum\limits_{k=1}^{k=G}{\xi (S_{vk}^{{}},{{G}_{vk}})} \cdot {B}\cdot {{\log }_{2}}{{(1+\frac{{{p}_{v}}\cdot {{h}_{v}}\cdot {{d}_{v}}^{-\partial }}{{{\sigma }^{2}}})}}\cdot PR_v}  {{V_v}\cdot G}=C . \\ 
\end{equation}
By transferring parameters unrelated to vehicle $v$, we obtain:
\begin{equation}
G_{fair}^{v}=\frac{\sum\limits_{k=1}^{k=G}{\xi (S_{vk}^{{}},{{G}_{vk}})} \cdot {{\log }_{2}}{{(1+\frac{{{p}_{v}}\cdot {{h}_{v}}\cdot {{d}_{v}}^{-\partial }}{{{\sigma }^{2}}})}}\cdot PR_v}  {{V_v}}.
\label{eq18}	
\end{equation}
Thus, we derive the fairness index $G_{fair}^{v}$. The averaged network's fairness can be described as:

\begin{equation}
{{G}_{fair}}=\frac{\sum\limits_{v=1}^{{{N}_{v}}}{G_{fair}^{v}}}{{{N}_{v}}}.
\end{equation}

\section{Model of AoI}
\label{sec5}
Next, we proceed to analyze the  AoI within the RSU's communication range.
We define the communication between vehicle $v$ and the RSU as link $v$, resulting in $N_v$ total links in the network. The transmission model adopts an SHS framework for modeling \cite{aoi1,aoi2}.
\begin{figure}[bp]
	\centering
	\includegraphics[width=\linewidth, scale=1.00]{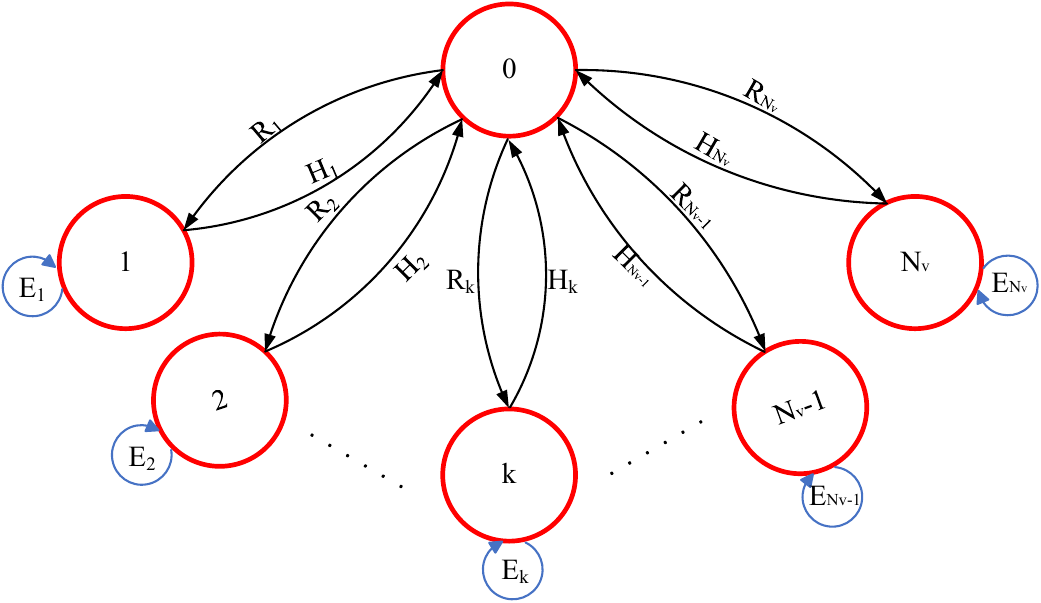}
	\caption{Markov Model}
	\label{fig3}
	\vspace{-0.0cm}
\end{figure}

Let $S(t)$ denote the current link state and $\boldsymbol{A(t)}$ represent the set of current AoI. Thus, the complete system model is formulated as $(S(t), \boldsymbol{A(t)})$,
$S(t)\in \{0,1,2,\ldots v,\ldots ,{N_v}\}$, $\boldsymbol{A(t)}=[A_r(t),A_v(t)]$.
When $ S(t) = 0$, it indicates the current link is idle (no vehicle communicating with RSU). When $s(t) = v$, it means vehicle $v$ is exchanging data with the RSU.	
$A_v(t)$ represents the AoI of link $v$ at time $t$, and $A_r(t)$ represents the RSU's AoI.  
After network setup is complete, $A_r(t)$ increases linearly with a slope of one, meaning the AoI at the RSU grows linearly over time. When the RSU receives data from vehicle $v$, it replaces $A_r(t)$ with the AoI from link $v$. Meanwhile, the AoI at link $v$, $A_v(t)$, begins linear growth with a slope of one after the vehicle generates a data packet. 
Therefore, we can model $S(t)$ as a discrete process and $\boldsymbol{A(t)}$ as a continuous process.  
Furthermore, transitions in the discrete process cause resets in the continuous process. Moreover, we set $l$ as a transition in the discrete process $S(t)$, and let $\boldsymbol{{{\psi }_{{{l}}}}}$ be the transition mapping matrix for $l$. 

Let $\mathbf{A}'$ and $\mathbf{A}$ represent the state of AoI through transition $l$. Similarly, we define $S_l$ and $S_l'$ to denote the discrete process before and after transition $l$, while $\eta_l$ represents the transition rate of $l$. This can be further categorized into three types of transition rates:
\begin{itemize}
\item[1)] Average transmission failure rate $R_v$, representing the transition rate caused by interference or packet collisions;
\item[2)]  Average service rate $H_v$, denoting the average transition rate for successful packet transmissions;
\item[3)] Average re-evaluation rate $E_v$, introduced for vehicle retransmission mechanisms.
\end{itemize}

Specifically, when transmission fails, the system transitions from state $0$ to state $v$ at rate $R_v$ due to retransmission.
Upon successful transmission, the system transitions from state $v$ back to idle state $0$ at rate $H_v$.
During re-evaluation, the system remains in state $v$ but triggers a re-evaluation at rate $E_v$.

We define $\boldsymbol{{\bar{Q}}}_{sl}=[{\bar{Q}}_{s0},Q_{s1}]$, where ${\bar{Q}}_{s0}$ as the steady-state correlation between $s_l$ and $A_r$, and ${\bar{Q}}_{s1}$ for $s_l$ and $A_v$. Accounting for continuous-process resets from discrete transitions: $\boldsymbol{{\bar{Q}}}' = \boldsymbol{{\bar{Q}}}_{s_l'}A'$,
where $\boldsymbol{{\bar{Q}}}'$ captures post-transition steady-state correlation. The SHS transition process is summarized in Table \ref{tab2}, which classifies transitions into three types:
1. Successful transmission.
2. Failed transmission.
3. Re-evaluation mechanism.
\begin{table}[t]
	\centering
	\resizebox{\columnwidth}{!}{%
		\begin{tabular}{cccccc}
			\hline
			\boldsymbol{$l$} & \boldsymbol{$s_l \rightarrow s_l'$} & \boldsymbol{$\eta_l$} & \boldsymbol{$S' = S{{\psi }_{l}}$} & \boldsymbol{${{\psi }_{l}}$} & \boldsymbol{$\overline{Q}_{S_l'} = \overline{Q}_{S_l} {{\psi }_{l}}$} \\
			\hline
			$1$ & $0 \rightarrow 1$ & $R_1$ & $[A_r, A_v]$ & $\begin{bmatrix}1 & 0 \\ 0 & 1\end{bmatrix}$ & $[\overline{Q}_{00}, \overline{Q}_{01}]$ \\
			$\vdots$ & $\vdots$ & $\vdots$ & $\vdots$ & $\vdots$ & $\vdots$ \\
			$N$ & $0 \rightarrow N$ & $R_{N_v}$ & $[A_r, A_v]$ & $\begin{bmatrix}1 & 0 \\ 0 & 1\end{bmatrix}$ & $[\overline{Q}_{00}, \overline{Q}_{01}]$ \\
			\hline
			$N + 1$ & $1 \rightarrow 0$ & $H_1$ & $[A_r, A_v]$ & $\begin{bmatrix}0 & 1 \\ 1 & 0\end{bmatrix}$ & $[\overline{Q}_{10}, \overline{Q}_{11}]$ \\
			$\vdots$ & $\vdots$ & $\vdots$ & $\vdots$ & $\vdots$ & $\vdots$ \\
			$N + k$ & $k \rightarrow 0$ & $H_k$ & $[A_v, 0]$ & $\begin{bmatrix}0 & 1 \\ 0 & 0\end{bmatrix}$ & $[\overline{Q}_{k1}, 0]$ \\
			$\vdots$ & $\vdots$ & $\vdots$ & $\vdots$ & $\vdots$ & $\vdots$ \\
			$2N$ & $N \rightarrow 0$ & $H_{N_v}$ & $[A_r, A_v]$ & $\begin{bmatrix}1 & 0 \\ 1 & 1\end{bmatrix}$ & $[\overline{Q}_{N0}, \overline{Q}_{N1}]$ \\
			\hline
			\multicolumn{1}{c}{$2N+1$} & $1 \rightarrow 1$ & $E_{1}$ & $[A_r, A_v]$ & $\begin{bmatrix}1 & 0 \\ 0 & 1\end{bmatrix}$ & $[\overline{Q}_{10}, \overline{Q}_{11}]$ \\
			
			$\vdots$ & $\vdots$ & $\vdots$ & $\vdots$ & $\vdots$ & $\vdots$ \\
			$3N$& $N \rightarrow N$ & $E_{N_v}$ & $[A_r, A_v]$ & $\begin{bmatrix}1 & 0 \\ 0 & 1\end{bmatrix}$ & $[\overline{Q}_{10}, \overline{Q}_{11}]$ \\
			\hline
		
		\end{tabular}%
	}
	\caption{SHS Transitions model }
	\label{tab2}
\end{table}
Table \ref{tab2} lists the transitions:
\begin{itemize}
	
\item[1)] Transition $l_1$ represents the system state changing from 0 to other states. Taking link $v$ as an example, the idle channel being occupied by link $v$ indicates the previous transmission of link $v$ failed and requires retransmission at rate $R_v$. Since the transmission failed, the AOI at the RSU remains unchanged and continues its linear growth with slope 1. This transition can be expressed as:
\[
\boldsymbol{A}'=\boldsymbol{A}\cdot \boldsymbol{{{\psi }_{{{l}_{1}}}}}=[{{A}_{r}},{{A}_{v}}],{\boldsymbol{{\bar{Q}}}_{s{{l}_{1}}}}^{\prime }={\boldsymbol{{\bar{Q}}}_{s{{l}_{1}}}}\cdot \boldsymbol{{{\psi }_{{{l}_{1}}}}}=[{{\bar{Q}}_{k1}},0].
\]
\item[2)] ${{l}_{2}}=\{{{N}_{v}}+1,{{N}_{v}}+2,{{N}_{v}}+3,\ldots ,2{{N}_{v}}\}$. The transition $l_2$ is the change of the system state from other states to 0. Taking link $N_v + k$ as an example, this means that a previously busy channel becomes idle. This indicates that the last transmission on link $N_v + k$ was successful, so no further transmission is required. This transition occurs at a rate of $H_k$. Meanwhile, since the transmission was successful, the AoI at the RSU is reset to the AoI of link $N_v + k$, and the AoI of link $N_v + k$ is set to zero. Therefore, for this successful transmission, we have:
\[
\boldsymbol{A}'=\boldsymbol{A}\cdot \boldsymbol{{{\psi }_{{{l}_{2}}}}}=[{{A}_{v}},0], {\boldsymbol{{\bar{Q}}}_{s{{l}_{2}}}}^{\prime }={\boldsymbol{{\bar{Q}}}_{s{{l}_{2}}}}\cdot \boldsymbol{{{\psi }_{{{l}_{2}}}}}=[{{\bar{Q}}_{k1}},0]
\]  
For links other than $N_v + k$, the success of any link has no impact on link $k$'s AoI, that is,  
\[
\boldsymbol{A}'=\boldsymbol{A}\cdot \boldsymbol{{{\psi }_{{{l}_{2}}}}}=\boldsymbol{A},{\boldsymbol{{\bar{Q}}}_{s{{l}_{2}}}}^{\prime }={\boldsymbol{{\bar{Q}}}_{s{{l}_{2}}}}\cdot \boldsymbol{{{\psi }_{{{l}_{2}}}}}={\boldsymbol{{\bar{Q}}}_{s{{l}_{2}}}}
\]
\item[3)] Transition ${{l}_{3}}=\{2{{N}_{v}}+1,2{{N}_{v}}+2,2{{N}_{v}}+3,\ldots ,3{{N}_{v}}\}$, The transition $l_3$ is the transition of the system state from any state other than 0 back to itself. This means that the transmission process has been re-evaluated, but since the transmitted information, sender, and receiver have not changed, only the transmission time and communication resources have been updated. Taking link $v$ as an example, the transition rate is $E_v$. Therefore, the channel link state remains unchanged, and the AoI at the RSU as well as the AoI on the link do not undergo any other change, continuing their linear growth. We have:
\[
\boldsymbol{A}'=\boldsymbol{A}\cdot \boldsymbol{{{\psi }_{{{l}_{3}}}}}=[{{A}_{r}},{{A}_{v}}],{\boldsymbol{{\bar{Q}}}_{s{{l}_{3}}}}^{\prime }={\boldsymbol{{\bar{Q}}}_{s{{l}_{3}}}}\cdot \boldsymbol{{{\psi }_{{{l}_{3}}}}}=[{{\bar{Q}}_{00}},{{\bar{Q}}_{01}}].
\]  
\end{itemize}
Based on \cite{aoi1}, we know that the link $k$'s AoI is:
\begin{equation} 
	\bar{\Delta}_k = \sum\bar{Q}_{s0}, \quad \forall k \in {1,2,\dots,N_v}.
	\label{22}
\end{equation}
According to Eq. (\ref{22}), our next objective is to obtain ${{\bar{Q}}_{s0}}$.
First, according to the theory in \cite{aoi2}, we can obtain:
\begin{equation}
\begin{aligned}
  \boldsymbol{{{\bar{Q}}}_{s{{l}_{a}}}}(\sum\limits_{{{l}_{a}}}{{{\eta }_{{{l}_{a}}}}})=\boldsymbol{{{b}_{s}}}{{{\bar{\pi }}}_{s}}+\sum\limits_{{{l}_{2}}}{{{\eta }_{{{l}_{a}}}}}\boldsymbol{{{\bar{Q}}}_{s{{l}_{b}}}}\boldsymbol{{{\psi }_{{{l}_{b}}}}}, & {{l}_{a}}\in {{L}_{s}}, \\ 
 & {{l}_{b}}\in {{{{L}'}}_{s}},  
\end{aligned}
	\label{23}
\end{equation}
where $\boldsymbol{b_s}$ is a binary differential vector coefficient used to characterize the AoI under state $s$, and ${{\bar{\pi }}_{s}}$ is the steady-state probability of state $s$. Furthermore, $l_a$ and $l_b$  denote the transitions between discrete state inputs and outputs, while $L$ and $L'$ represent the sets of inputs and outputs transitions.

Considering that the vehicle’s data is generated only after it obtains communication resources, i.e., after it begin to transmit, the link $k$'s AoI increases linearly with slope one only when $s = k$. However, the AoI at the RSU keeps increasing until it receives data from the link. Therefore, $\boldsymbol{b_s}$ can be expressed as:
\begin{equation}
\left\{ \begin{aligned}
  & \boldsymbol{{{b}_{s}}}=[1,0],for\text{ }\forall s\ne k \\ 
 & \boldsymbol{{{b}_{s}}}=[1,1],for\text{ }\forall s=k. \\ 
\end{aligned} \right\}
\end{equation}

Next, we need to obtain ${{\bar{\pi }}_{s}}$. As stated by \cite{aoi2}, the ${{\bar{\pi }}_{s}}$ satisfies the following conditions:
\begin{equation}
  {{{\bar{\pi }}}_{s}}\sum\limits_{l\in {{L}_{s}}}{{{\eta }_{l}}}=\sum\limits_{l\in {{{{L}'}}_{s}}}{{{\eta }_{l}}}{{{\bar{\pi }}}_{sl}},\sum\limits_{s\in S}{{{{\bar{\pi }}}_{s}}}=1  ,\text{    s}\in S. 
\label{25}
\end{equation}
Substituting ${{\eta }_{l}}$ into Eq. (\ref{25}), the ${{\bar{\pi }}_{s}}$ is shown as:

\begin{equation}
\left\{ \begin{aligned}
  & {{{\bar{\pi }}}_{0}}=\frac{1}{{{N}_{F}}} \\ 
 & {{{\bar{\pi }}}_{k}}=\frac{{{R}_{k}}}{{{N}_{F}}({{H}_{k}}-{{E}_{k}})}, \\ 
\end{aligned} \right.
\end{equation}
where ${{N}_{F}}$ is a normalization factor, which is:
\begin{equation}
{{N}_{F}}=1+\sum\limits_{k=1}^{N_v}{\frac{{{R}_{k}}}{{{H}_{k}}-{{E}_{k}}}}.
\end{equation}
Next, when $s=0$, according to the analysis in \cite{wan}, we know that the left side of Eq. (\ref{23}) represents the system state transitioning to busy—that is, transitioning to a state not equal to zero—while the right side represents the system state transitioning to idle, i.e., $s$ transitions to 0. When $s=0$, the channel is idle, so $\boldsymbol{{{b}_{s}}} = [1,0]$. The transitions on the left side correspond to states from 1 to $n$ and from $2n+1$ to $3n$, while the right side correspond to states from $n+1$ to $2n$.

Moreover, based on the analysis in \cite{aoi2}, Eq. (\ref{23}) applies to any reset mapping and therefore also applies to mappings caused by the re-evaluation mechanism. Substituting $s=0$ into Eq. (\ref{23}), we obtain:
\begin{equation}
{{\bar{Q}}_{00}}(\sum\limits_{s=1}^{{{N}_{v}}}{{{R}_{s}}})+\sum\limits_{\begin{smallmatrix} 
 s=1 \\ 
 s\ne k 
\end{smallmatrix}}^{{{N}_{v}}}{{{{\bar{Q}}}_{s0}}}\cdot {{E}_{s}}={{\bar{\pi }}_{0}}+\sum\limits_{\begin{smallmatrix} 
 s=1 \\ 
 s\ne k 
\end{smallmatrix}}^{{{N}_{v}}}{{{H}_{s}}}{{\bar{Q}}_{s0}}+{{H}_{k}}{{\bar{Q}}_{k1}},
\label{28}
\end{equation}

\begin{equation}
{{\bar{Q}}_{01}}(\sum\limits_{s=1}^{{{N}_{v}}}{{{R}_{s}}})+\sum\limits_{\begin{smallmatrix} 
 s=1 \\ 
 s\ne k 
\end{smallmatrix}}^{{{N}_{v}}}{{{{\bar{Q}}}_{s1}}}\cdot {{E}_{s}}=\sum\limits_{\begin{smallmatrix} 
 s=1 \\ 
 s\ne k 
\end{smallmatrix}}^{{{N}_{v}}}{{{H}_{s}}}{{\bar{Q}}_{s1}}.
\label{29}
\end{equation}

Next, considering the case when $s \neq 0$, the left side of Eq. (\ref{23}) represents the system state transitioning to idle, while the right side represents the system state transitioning to busy, i.e., transitioning to a state not equal to 0. The transitions on the left correspond to states from $n+1$ to $2n$, and those on the right denote states from 1 to $n$ and from $2n+1$ to $3n$. Substituting $s \neq 0$ into Eq. (\ref{23}), we obtain:

\begin{equation}
{{\bar{Q}}_{s0}}\cdot {{H}_{s}}={{\bar{\pi }}_{s}}+{{R}_{s}}\cdot {{\bar{Q}}_{00}}+{{E}_{s}}{{\bar{Q}}_{s0}},for\text{ }\forall s=\{1,2,...,{{N}_{v}}\}
\label{30}
\end{equation}

\begin{equation}
{{\bar{Q}}_{s1}}\cdot {{H}_{s}}={{R}_{s}}\cdot {{\bar{Q}}_{01}}+{{\bar{Q}}_{s1}}{{E}_{s}},for\text{ s}\ne k
\label{31}
\end{equation}

\begin{equation}
{{\bar{Q}}_{k1}}\cdot {{H}_{k}}={{\bar{\pi }}_{k}}+{{R}_{k}}\cdot {{\bar{Q}}_{01}}+{{\bar{Q}}_{k1}}{{E}_{k}},for\text{ }s=k.
\label{32}
\end{equation}

Next, based on Eq. (\ref{28}) to Eq. (\ref{32}), we proceed to derive ${\bar{Q}}_{s0}$.
First, according to Eq. \ref{30}, we have:
\begin{equation}
{{\bar{Q}}_{s0}}=\frac{{{{\bar{\pi }}}_{s}}+{{R}_{s}}\cdot {{{\bar{Q}}}_{00}}}{{{H}_{s}}-{{E}_{s}}}.
\label{33}
\end{equation}

According to Eq. (\ref{33}), it can be seen that the next step is to derive $ {{{\bar{Q}}}_{00}}$.
To obtain $ {{{\bar{Q}}}_{00}}$, we note from Eq. (\ref{28}) that it requires ${{{\bar{Q}}}_{k1}}$. We rearrange Eq. \ref{31}:

\begin{equation}
{{\bar{Q}}_{s1}}=\frac{{{R}_{s}}\cdot {{{\bar{Q}}}_{01}}}{{{H}_{s}}-{{E}_{s}}}.
\label{34}
\end{equation}
We rearrange Eq. (\ref{32}) as follows:

\begin{equation}
{{\bar{Q}}_{k1}}=\frac{{{{\bar{\pi }}}_{k}}+{{R}_{k}}\cdot {{{\bar{Q}}}_{01}}}{{{H}_{k}}-{{E}_{k}}}.
\label{35}
\end{equation}
Substituting Eq. (\ref{34}) into Eq. (\ref{29}), we obtain:

\begin{equation}
{{\bar{Q}}_{01}}\cdot {{R}_{k}}\cdot (1+\frac{{{E}_{k}}}{{{H}_{k}}-{{E}_{k}}})=0.
\label{36}
\end{equation}
Since $R_K$, $E_K$, and $H_K$ are all greater than zero, We get:
\begin{equation}
{{\bar{Q}}_{01}}=0.
\label{37}
\end{equation}
Therefore, substituting Eq. (\ref{37}) into Eq. (\ref{35}), we obtain:
\begin{equation}
{{\bar{Q}}_{k1}}=\frac{{{{\bar{\pi }}}_{k}}}{{{H}_{k}}-{{E}_{k}}}.
\label{38}
\end{equation}
Finally, substituting Eq. (\ref{33}) and Eq. (\ref{38}) into Eq. (\ref{28}), we obtain:
\begin{equation}
{{\bar{Q}}_{00}}=\frac{{{H}_{k}}-{{E}_{k}}}{{{H}_{k}}\cdot {{R}_{k}}}.
\label{39}
\end{equation}

Then, according to Eq. (\ref{33}), we obtain ${\bar{Q}}_{s0}$. In the end, we substitute Eq. (\ref{39}) and Eq. (\ref{33}) into Eq. (\ref{22}), the AoI of link $k$ can be expressed as:
\begin{equation}
\begin{aligned}
  & {{{\bar{\Delta }}}_{k}}=\sum\limits_{s=0}^{{{N}_{v}}}{{{{\bar{Q}}}_{s0}}} \\ 
 & =\frac{{{H}_{k}}-{{E}_{k}}}{{{H}_{k}}\cdot {{R}_{k}}}[1+\sum\limits_{s=1}^{{{N}_{v}}}{\frac{{{R}_{s}}}{{{H}_{s}}-{{E}_{s}}}}]+\sum\limits_{s=1}^{{{N}_{v}}}{\frac{{{{\bar{\pi }}}_{s}}}{{{H}_{s}}-{{E}_{s}}}}  .
\end{aligned}
\label{40}
\end{equation}
Subsequently, we get the mean AoI:
\begin{equation}
\bar{\Delta }=\frac{\sum\limits_{k=1}^{{{N}_{v}}}{{{{\bar{\Delta }}}_{k}}}}{{{N}_{v}}},
\label{41}
\end{equation}
where $\bar{\Delta }$ denotes the network's averaged AoI.

Finally, we define the transfer rates under NR V2X Mode 2, namely \( H_v \), \( R_v \), and \( E_v \).
According to \cite{wan}, the average service rate can be expressed as:
\begin{equation}
{{H}_{v}}=\frac{1}{{{T}_{v}^H}},for\text{ }\forall i\in \{1,2,....N_v\},
	\label{42}
\end{equation}
where ${T}_{v}^H$ represents the average time for successful transmission for vehicle $v$,  which can be expressed as \cite{latency}:
\begin{equation}
{T}_{v}^H={{t}_{p}^v}+{{t}_{fa}^v}+{{t}_{w}^v}+{{t}_{t}^v},
	\label{43}
\end{equation}
where ${{t}_{p}^v}$ is the data processing time, ${{t}_{fa}^v}$ is the frame alignment duration, ${{t}_{w}^v}$ the resource allocation period i.e., the vehicle’s selection window size, and ${{t}_{t}^v}$ the data transmission time. Likewise, the average failure rate can be represented by:
\begin{equation}
{{R}_{v}}=\frac{1}{{{T}_{v}^R}},for\text{ }\forall i\in \{1,2,....N_v\},
	\label{44}
\end{equation}
\begin{equation}
T_{v}^{R}=T_{v}^{H}+n\cdot {{T}_{r}},
\label{45}
\end{equation}
where $T_r$ denotes the retransmission time after a failure.
Finally, the re-evaluation mechanism transition rate is:
\begin{equation}
{{E}_{v}}=\frac{1}{T_{v}^{E}},
\label{46}
\end{equation}
where $T_{v}^E$ denotes the average time of the re-evaluation, which is autonomously determined by the vehicle \cite{5G}.

\section{Optimization Objective and  Solutions }
\label{sec6}
In this part, by applying the sequential convex approximation \cite{SCA}, the non-convex problem is reformulated into a convex approximation. Meanwhile, considering that convexifying a non-convex problem requires complicated manual derivations, we also take into account solving it using a LLM-based MOEA/D algorithm (LLM-MO) \cite{LLM}.
\subsection{Optimization Objective}

Based on section \ref{sec4}, we can define the fair access problem as a multi-objective optimization problem:
\begin{equation}
	\begin{aligned}
		G_v(\boldsymbol{w}) =  \left| G_{fair}^{v}(\boldsymbol{w}) - G_{fair}(\boldsymbol{w}) \right|, \\
		 s.t \text{    }\boldsymbol{w} = \{w^1, w^2, \dots, w^{N}_{v}\}, \\
		v \in \{1,2,\dots,N_v\}.
	\end{aligned}
	\label{47}
\end{equation}
Meanwhile, consider the AoI:
$G_{{N}_{v}+1}(\boldsymbol{w})=\min \bar{\Delta }$.
Therefore, the overall optimization objective is given by:
\begin{equation}
\begin{aligned}
   \min_{\boldsymbol{w}} \ \boldsymbol{G(w)}&=[G_1(\boldsymbol{w}), G_2(\boldsymbol{w}), \dots, G_{{N}_{v}}(\boldsymbol{w}),  G_{{N}_{v}+1}(\boldsymbol{w})]^T\\ 
  s.t \text{    } C_1&:{{V}_{\text{min}}}\le {{V}_{v}}\le {{V}_{\text{max}}}, v \in [1, \ldots, {N}_{v}], \\ 
 \text{           }  C_2&: {{w}_{\min }}\le {{w}_{v}}\le {{w}_{\max }}, v \in [1, \ldots, {N}_{v}]. \\ 
\end{aligned}
\label{48}
\end{equation}
According to the 3GPP standard \cite{3gpp}, ${w}_{\min }$ and ${w}_{\max }$ denote the boundary limitation of windows.

\subsection{SCA Algorithm}

Since Eq. (\ref{48}) is a multi-objective optimization problem, we first need to transform it into a single-objective problem. The problem contains several non-convex components, such as absolute value terms, $P{{R}_{v}}$ and ${{H}_{v}}\cdot {{R}_{v}}$.  To deal with this problem, we consider the auxiliary variables to replace the non-convex terms and use Taylor expansion to approximate the non-convex parts of the objective and constraints. The final problem is transformed into a convex problem, which is able to be efficiently addressed through classical convex optimization, including interior-point algorithms and Alternating Direction Method of Multipliers (ADMM) \cite{ADMM}.

First, we use the weighted sum method \cite{weight} to convert the original optimization problem into a single-objective problem:
\begin{equation}
\begin{aligned}
   \min_{\boldsymbol{w}} \ \boldsymbol{{G(w)}} &= \sum\nolimits_{v=1}^{{{N}_{v}}}{{{\lambda }_{v}}\cdot G_v(\boldsymbol{w}) }+{{\lambda }_{{{N}_{v}}+1}}\cdot G_{{N}_{v}+1}(\boldsymbol{w}),\\  
   s.t\text{    } &C_1, C_2, 
   \label{49}
\end{aligned}
\end{equation}
where the ${\lambda }_{v}, v \in [1, \ldots, {N}_{v}+1]$ is the weight used to keep the balance between different objectives.
Since $\left| G_{fair}^{v}(\boldsymbol{w}) - G_{fair}(\boldsymbol{w}) \right|$ is a non-convex term, in order to remove the absolute value term, we introduce an auxiliary variable ${{z}_{v}}\ge 0$ which satisfies:
\begin{equation}
\begin{aligned}
  & C_3:{{z}_{v}}\ge G_{fair}^{v}(\boldsymbol{w}) - G_{fair}(\boldsymbol{w}), \\ 
 & C_4:{{z}_{v}}\ge -(G_{fair}^{v}(\boldsymbol{w}) - G_{fair}(\boldsymbol{w})). \\ 
\end{aligned}
\end{equation}
Thus, Eq. (\ref{49}) can be rewritten as:
\begin{equation}
\begin{aligned}
   \min_{\boldsymbol{w}} \ \boldsymbol{{G(w)}}&=\sum\nolimits_{v=1}^{{{N}_{v}}}{{{\lambda }_{v}}{{z}_{v}}}+{{\lambda }_{{{N}_{v}}+1}}\cdot G_{{N}_{v}+1}(\boldsymbol{w}) \\ 
 s.t \text{    } & C_1, C_2, C_3, C_4.
    \label{51}
\end{aligned}
\end{equation}
However, since there are still non-convex terms in $G_{fair}^{v}(\boldsymbol{w})$ and $G_{{N}_{v}+1}(\boldsymbol{w})$, Eq. (\ref{51}) remains a non-convex function, we choose to apply the Sequential Convex Approximation (SCA) method to handle the non-convex terms. 
First, we tansfer $G_{fair}^{v}(\boldsymbol{w})$ into convex.
Since the non-convex terms in $G_{fair}^{v}(\boldsymbol{w})$ is $P{{R}_{v}}$ which involves extensive products, it will leads to an increase in complexity. So we apply the logarithm to transform the product terms into a summation form:
\begin{equation}
\log P{{R}_{v}}=\underset{j\ne v}{\mathop{\sum }}\,\log \left( 1-{{\delta }_{v,j}(w)} \right).
    \label{52}
\end{equation}
Note that the problem is still non-convex because $\delta_{v,j}$ is non-convex, and $\log(1 - x)$ is a concave function. For each term, we perform SCA at the current $t$-th iteration point $\delta_{v,j}^{(t)}$. By applying a Taylor expansion to each term, we get:
\begin{equation}
\log (1-{{\delta }_{v,j}(w)})= \log (1-\delta _{v,j}^{(t)})-\frac{1}{1-\delta _{v,j}^{(t)}}\cdot \left( {{\delta }_{v,j}}(w)-\delta _{v,j}^{(t)} \right).
    \label{53}
\end{equation}
Substituting this into the previous expression, we obtain:
\begin{equation}
\log P{{R}_{v}}= C_{v}^{(t)}-\underset{j\ne v}{\mathop{\sum }}\,\frac{1}{1-\delta _{v,j}^{(t)}}\cdot \left( {{\delta }_{v,j}}(w)-\delta _{v,j}^{(t)} \right),
    \label{54}
\end{equation}
where $C_{v}^{(t)}=\underset{j\ne v}{\mathop{\sum }}\,\log (1-\delta _{v,j}^{(t)})$ is a constant.
We continue by substituting the explicit form of $\delta_{v,j}(w)$ into the above approximated expression, resulting in a convex approximation formula that is more suitable for numerical solving. We then perform a first-order Taylor expansion on $\delta_{v,j}(w)$:
\begin{equation}
{{\delta }_{v,j}}(w)={{\delta }_{v,j}}({{w}^{(t)}})+{{\nabla }_{w}}{{\delta }_{v,j}}{{({{w}^{(t)}})}^{\top }}(w-{{w}^{(t)}}).
    \label{55}
\end{equation}
We combine the terms that are not optimization variables into $c$. Since $\delta_{v,j}(w)$ is a function of $w_v$ and $w_j$, we take the partial derivatives with respect to $w_v$ and $w_j$, yielding:
\begin{equation}
\frac{\partial {{\delta }_{v,j}}}{\partial {{w}_{v}}}=\frac{\partial }{\partial {{w}_{v}}}\left( \frac{{{[({{w}_{v}}+1)({{w}_{j}}+1)]}^{2}}}{{{c}}({{w}_{v}}+{{w}_{j}}+1)} \right),
    \label{56}
\end{equation}
which can be numerically computed in each iteration. We now summarize the previous results as follows:
\begin{equation}
\begin{aligned}
&\log P{{R}_{v}}(w)= {\mathop{\underset{{}}{\mathop{\underset{j\ne v}{\mathop{\sum }}\,\log (1-\delta _{v,j}^{(t)})}}\,}}\,-\underset{j\ne v}{\mathop{\sum }}\,\frac{1}{1-\delta _{v,j}^{(t)}}\cdot\\
& \left[ {{\nabla }_{{{w}_{v}}}}\delta _{v,j}^{(t)}({{w}_{v}}-w_{v}^{(t)})+{{\nabla }_{{{w}_{j}}}}\delta _{v,j}^{(t)}({{w}_{j}}-w_{j}^{(t)}) \right].
\end{aligned}
    \label{57}
\end{equation}
After organizing, we obtain:
\begin{equation}
\log P{{R}_{v}}= C_{v}^{(t)}+\underset{j\ne v}{\mathop{\sum }}\,a_{v,j}^{(t)}({{w}_{v}}-w_{v}^{(t)})+b_{v,j}^{(t)}({{w}_{j}}-w_{j}^{(t)}).
    \label{58}
\end{equation}
where 
\begin{equation}
a_{v,j}^{(t)}=-\frac{1}{1-\delta _{v,j}^{(t)}}\cdot {{\left. \frac{\partial {{\delta }_{v,j}}}{\partial {{w}_{v}}} \right|}_{{{w}^{(t)}}}}b_{v,j}^{(t)}=-\frac{1}{1-\delta _{v,j}^{(t)}}\cdot {{\left. \frac{\partial {{\delta }_{v,j}}}{\partial {{w}_{j}}} \right|}_{{{w}^{(t)}}}}
    \label{59}
\end{equation}
We have transformed $G_{fair}^{v}(\boldsymbol{w})$ into a convex function:
\begin{equation}
\begin{aligned}
& G_{fair}^{v}(\boldsymbol{w})=\frac{\sum\limits_{k=1}^{k=G}{\xi (S_{vk}^{{}},{{G}_{vk}})} \cdot {{\log }_{2}}{{(1+\frac{{{p}_{v}}\cdot {{h}_{v}}\cdot {{d}_{v}}^{-\partial }}{{{\sigma }^{2}}})}}}  {{V_v}}\cdot \\
&\exp (-(C_{v}^{(t)}+\underset{j\ne v}{\mathop{\sum }}\,(a_{v,j}^{(t)}({{w}_{v}}-w_{v}^{(t)})+b_{v,j}^{(t)}({{w}_{j}}-w_{j}^{(t)})))).
\end{aligned}
    \label{60}
\end{equation}
Next, we consider $\bar{\Delta }(w)$.
We divide ${{\bar{\Delta }}}_k$ into two parts:
\begin{equation}
\begin{aligned}
  A({{w}_{k}}) &=\frac{{{H}_{k}}-{{E}_{k}}}{{{H}_{k}}\cdot {{R}_{k}}}[1+\sum\limits_{s=1}^{{{N}_{v}}}{\frac{{{R}_{s}}}{{{H}_{s}}-{{E}_{s}}}}] \\ 
 &\text{          }= \left(w_k+c_2^k \right) \left[ 1-c_3^k({{w}_{k}}+c_1^k) \right]\cdot  \\ 
  &\text{          }\left( 1+\underset{s=1}{\overset{{{N}_{v}}}{\mathop{\sum }}}\,\frac{{{w}_{s}}+{{c}_{1}}}{({{w}_{s}}+c_2^s)\left[ 1-c_3^s({{w}_{s}}+{c}_{1}^{s}) \right]} \right) \\ 
\end{aligned}
    \label{61}
\end{equation}
\begin{equation}
B({\boldsymbol{w}})=\sum\limits_{s=1}^{{{N}_{v}}}{\frac{{{{\bar{\pi }}}_{s}}}{{{H}_{s}}-{{E}_{s}}}},
    \label{62}
\end{equation}
where $c_1^k, c_2^k$ and $c_3^k$ respectively denote the terms independent of $w$ in $H_k,R_k$ and $E_k$. 
We proceed with a Taylor expansion, which first requires taking derivatives.
To facilitate differentiation, we have:
$A({{w}_{k}})={{A}_{1}}({{w}_{k}})\cdot {{A}_{2}}({{w}_{k}})$, where:
\begin{equation}
\begin{aligned}
  & {{A}_{1}}(w)=({{w}_{k}}+{{c}_{2}^k})\left[ 1-{{c}_{3}^k}({{w}_{k}}+{{c}_{1}^k}) \right]. \\ 
 & {{A}_{2}}(w)=\left( 1+\underset{s=1}{\overset{{{N}_{v}}}{\mathop{\sum }}}\,\frac{{{w}_{s}}+{{c}_{1}^s}}{({{w}_{s}}+{{c}_{2}^s})\left[ 1-{{c}_{3}^s}({{w}_{s}}+{c}_{1}^{s}) \right]} \right). \\ 
\end{aligned}
    \label{63}
\end{equation}
Therefore, we have:
\begin{equation}
\frac{\partial A({{w}_{k}})}{\partial {{w}_{k}}}=\frac{d{{A}_{1}}({{w}_{k}})}{d{{w}_{k}}}\cdot {{A}_{2}}({{w}_{k}})+{{A}_{1}}({{w}_{k}})\cdot \frac{d{{A}_{2}}({{w}_{k}})}{d{{w}_{k}}} .
    \label{64}
\end{equation}
According to the first-order Taylor expansion, we have:
\begin{equation}
A({{w}_{k}})= A({{w}_{k}}^{(t)})+{A}'({{w}_{k}}^{(t)})(w-w_{k}^{(t)}).
    \label{65}
\end{equation}
Next, we apply the same method to process $B({\boldsymbol{w}})$:
\begin{equation}
B({\boldsymbol{w}})= \underset{s}{\mathop{\sum }}\,\left[ B(w_{s}^{(t)})+\nabla B(w_{s}^{(t)})({{w}_{s}}-w_{s}^{(t)}) \right].
    \label{66}
\end{equation}
Thus, convex approximation of ${{\bar{\Delta }}_{k}}$ can be expressed as:
\begin{equation}
{{{\bar{\Delta }}}_{k}}={A}(w_k)+{B}({\boldsymbol{w}}).
    \label{67}
\end{equation}
The AoI is given by:
\begin{equation}
{\bar{\Delta }}({\boldsymbol{w}})=\frac{\sum\limits_{k=1}^{{{N}_{v}}}{{{{{\bar{\Delta }}}}_{k}}}}{{{N}_{v}}}.
    \label{68}
\end{equation}
Now, we have transformed the non-convex terms in the optimization objective into convex functions. Therefore, the optimization objective is given by:
\begin{equation}
\begin{aligned}
   \min_{\boldsymbol{w}} \ \boldsymbol{{G(w)}}&=\sum\nolimits_{v=1}^{{{N}_{v}}}{{{\lambda }_{v}}{{z}_{v}}}+{{\lambda }_{{{N}_{v}}+1}}{\bar{\Delta }}({\boldsymbol{w}}) \\ 
 & s.t\text{    }C_1,C_2,C_3,C_4. \\ 
\end{aligned}
    \label{69}
\end{equation}
Next, we can adopt convex optimization methods to solve this problem. In each SCA iteration, a new optimization problem is formulated and solved using ADMM to obtain the current solution $\boldsymbol{w^t}$. To prevent the $\boldsymbol{w}^{(t+1)}$ from spanning too widely compared with $\boldsymbol{w}^{(t)}$, we introduce $\beta$ as the learning rate to limit the update speed of $\boldsymbol{w}^{(t+1)}$. The iteration proceeds until either convergence or $G_{max}$ is achieved. The algorithm framework is shown in Algorithm \ref{alg1}.
\begin{algorithm}[tb]
	\caption{SCA Iteration Algorithm}
	\label{alg1}
    \KwInput{speed $V$, vehicle numbers $N_v$, tolerance $\varepsilon$, maximum iteration $G_{max}$, learning rate $\beta$.}
	\KwOutput{$\boldsymbol{w}^{*}$}
	
	\textbf{Initialization Phase:} \\
    Set iteration index $t = 0$ \\
    Initialize variables $\boldsymbol{w}^{(0)}$ \\
    Set convergence tolerance $\varepsilon$ and  $G_{max}$ \\
    
	\textbf{Main Iteration:} \\
    \While{$\|\boldsymbol{w}^{(t+1)} - \boldsymbol{w}^{(t)}\| \leq \varepsilon$ and $t < G$}{
        \textbf{Step 1:} Formulate the single-objective optimization problem based on Eq. \ref{49}.  \\
        \textbf{Step 2:} Introduce auxiliary variables to construct Eq. \ref{51}.  \\
        \textbf{Step 3:} Convexify $G_{fair}^{v}(\boldsymbol{w})$ with Eq. \ref{52}-\ref{60}.\\
        \textbf{Step 4:} Convexify $G_{N_v+1}(\boldsymbol{w})$ with Eq. \ref{61}-\ref{68}.\\
        \textbf{Step 5:} Formulate the current optimization problem \ref{69} based on the current solution $\boldsymbol{w}^{(t)}$.\\
        \textbf{Step 6:} Apply ADMM to solve \ref{69} to get $\boldsymbol{w}^{(t+1)}$.\\
        \textbf{Step 7:} Update $\boldsymbol{w}^{(t+1)}=\beta \cdot \boldsymbol{w}^{(t+1)}+(1-\beta)\cdot \boldsymbol{w}^{(t)} $.\\
        \textbf{Step 8:} Check for convergence:
        \If{$\|\boldsymbol{w}^{(k+1)} - \boldsymbol{w}^{(t)}\| \leq \varepsilon$}{
            \textbf{Exit the loop.} \\
        }
        Increment $t$ \\
    }
    \KwRet{$\boldsymbol{w}^{*} = \boldsymbol{w}^{(t+1)}$}
\end{algorithm}
\subsection{LLM-MO Algorithm}

The proposed LLM-MO algorithm builds upon the traditional MOEA/D framework \cite{LLM}, with the key novelty being the integration of an LLM-guided crossover mechanism. The overall process can be divided into three main stages: initialization, iteration, and solution update.

In the initialization phase:
$\boldsymbol{m}$ weight vectors are generated using the Das–Dennis uniform sampling method. Each weight vector corresponds to a subproblem in MOEA/D.
For each weight vector, a neighborhood is defined by selecting the $T$ closest vectors in terms of Euclidean distance.
A population of candidate solutions is randomly initialized.
The ideal point is computed from the initialized population.

During each iteration, with probability $P_{near}$, parent solutions are selected from the subproblem’s neighborhood, or with probability $1-P_{near}$, from the entire population to promote diversity.
Unlike conventional crossover, offspring solutions are generated by an LLM via prompt engineering.
The prompt design must include: (i) a description of the optimization task, (ii) input data (the window size of the previous round and its fairness index difference and AoI), and (iii) the expected output format (better window size).
An example prompt is:

\begin{tcolorbox}[colframe=blue!50!black, colback=yellow!10, coltitle=black, 
                  sharp corners=southwest, boxrule=0.8mm]
You will perform a multi-objective optimization task. I will provide you with multiple sets of data, where each set includes window size and their corresponding fairness index differences and AoI. Each set starts with $<$begin$>$ and finishes with $<$end$>$. 

point: $<$begin$>$0.124,0.352,0.421 $<$end$>$

value: $<$begin$>$0.021,0.031,0.012,67 $<$end$>$

Based on these data, provide one set of window sizes different from the above data, whose objective values are better than the given data. Your answer should start with $<$begin$>$ and finish with $<$end$>$.
\end{tcolorbox}

Once offspring solutions are obtained:
(1) The ideal point is updated to the better fairness index difference and AoI.
(2) The neighborhood solutions are updated by replacing inferior $\boldsymbol{w}$ with better offspring solutions.
(3) This iterative replacement  improves the population quality and maintains diversity.

The algorithm repeats the above steps until  maximum iterations or convergence threshold is satisfied. At this point, the algorithm outputs the final Pareto front. (Steps 1 to 17.)

Finally, we will select the optimal solution from the Pareto solution set. After eliminating some unreasonable objective values, we choose the lowest AoI as the optimal one. (Steps 18 to 23 in algorithm \ref{alg2}.)
\begin{algorithm}[tb]
\caption{LLM-MO: LLM-guided MOEA/D }
\label{alg2}
\KwIn{Number of objectives $N_v+1$, number of subproblems $S$, neighborhood size $T$, 
       neighborhood selection probability $P_{\mathrm{near}}$, 
       maximum iterations $N_{\mathrm{iter}}$, vehicle speed  $V$}
\KwOut{$\boldsymbol{w}^*$}

Define weight $\{ \boldsymbol{m}^1,\dots,\boldsymbol{m}^S \}$ by Das-Dennis sampling\;
\For{each subproblem $s \in \{1,\dots,S\}$}{
    Define neighborhood $B^s$ with  $T$ nearest weight\;
}

Randomly initialize population $\mathcal{P}=\{\boldsymbol{w}^1,\dots,\boldsymbol{w}^S\}$\;
Evaluate $\boldsymbol{G}(\boldsymbol{w}^i) = (G_1,\dots,G_{N_v+1})$ for all $i \in S$\;
Initialize ideal point $z^* = \{\min{G_1},\dots,\min{G_{N_v+1}}\}$

\For{$iter \leftarrow 1$ \KwTo $N_{\mathrm{iter}}$}{
    \For{$s \leftarrow 1$ \KwTo $S$}{
        With probability $P_{\mathrm{near}}$ select parents from $B^s$; otherwise select from $\mathcal{P}$\;
        Construct LLM prompt with parents\;
        Query LLM to generate offspring $\boldsymbol{w_{new}}$\;
        Evaluate offspring $\boldsymbol{G(w_{new})}$\;
        Update ideal point:
        $
        \boldsymbol{z^*} \leftarrow \min\big(\boldsymbol{z^*}, \boldsymbol{G(w_{new})} \big)
        $
        
        Define the subproblem as:
        $g(\boldsymbol{w}\mid \boldsymbol{m}_s,\boldsymbol{z^*}) 
        = \max_{1\le n\le N_v+1}\ \boldsymbol{m}_s\,|\,G_n(\boldsymbol{w}) - z_n^*|
        $
        
        \For {each $j \in B^s$:}{
        \If{$g(\boldsymbol{w_{new}}\mid \boldsymbol{m}^j,\boldsymbol{z^*}) 
            \le g(\boldsymbol{w}^j\mid \boldsymbol{m}^j,\boldsymbol{z^*})$}{
            $\boldsymbol{w}^j \leftarrow \boldsymbol{w_{new}}$\;
        }
        }
    }
}

Set ${\bar{\Delta }}({\boldsymbol{w}}) = +\infty$\;

\ForEach{$p \in \mathcal{P}$}{
    \If{$ G_{i} \leq G_{max},i \in p$}{
        \If{$G_{N_v+1} \leq age$}{
            ${\bar{\Delta }}({\boldsymbol{w}}) \gets G_{N_v+1}$\;
            $\boldsymbol{w}^* \gets \boldsymbol{w}^p$\;
        }
    }
}

\Return{$\boldsymbol{w}^*$}
\end{algorithm}



\section{Numerical Simulation Result And Analysis}
\label{sec7}

\begin{table}[tb]
	\caption{Important parameters}
	\label{tab3}
	\centering
	\begin{tabular}{|c|c|c|c|}
		\hline
		\textbf{Parameters} &\textbf{Value} &\textbf{Parameters} &\textbf{Value}\\
		\hline
		$R$ & $200m$ &  $G_{max}$ & $5000$ \\
		\hline
		$B$ & $20MHZ$ & $\alpha$ & $3$\\
		\hline
		$w_{min}$ & $20ms$ & $w_{max}$ & $150ms$ \\
		\hline
        $\beta$ & $0.2$ & $RRI$ & $100ms$ \\
		\hline
		$N_v$ & $3$ & $\sigma^2$ &$ 9dB$ \\
		\hline
		$\varepsilon$ & $0.01$ & $t_{fa}$ & $0.468ms$ \\
		\hline
		$V_{min}$ & $20m/s$ & $V_{max}$ & $30m/s$ \\
		\hline
	\end{tabular}
\end{table}

We will next demonstrate the reliability of the proposed scheme through numerical simulations. The benchmark algorithms for this experiment are several multi-objective algorithms. Our baseline algorithms include  NSGA-II, SPEA2, NSGA-III, and MOEA/D \cite{moea1}. The baseline multi-objective algorithms are realized via the pymoo framework \cite{pymoo}. Meanwhile, this paper uses the QSOP solver in the cvxpy framework for convex optimization, which adopts ADMM \cite{cvxpy}. The parameter settings are shown in Table \ref{tab3}.

\begin{figure}[htbp]
	\centering
	\includegraphics[width=\linewidth, scale=0.9]{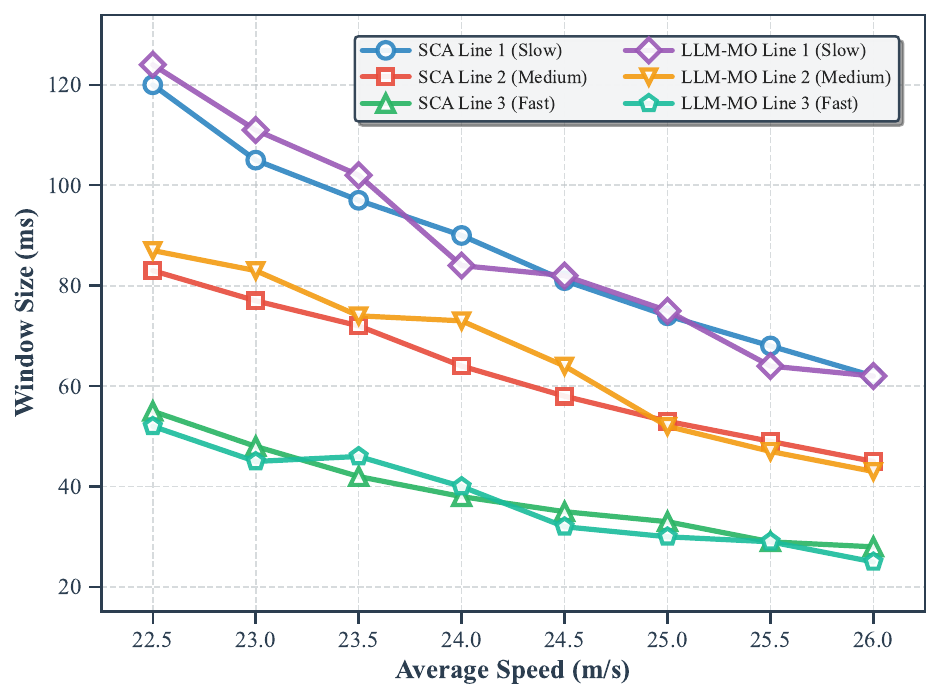}
	\caption{Different Lines' Optimal Selection Window}
	\label{fig4}	
\end{figure}
\begin{figure}[htbp]
	\centering
	\includegraphics[width=\linewidth, scale=1.00]{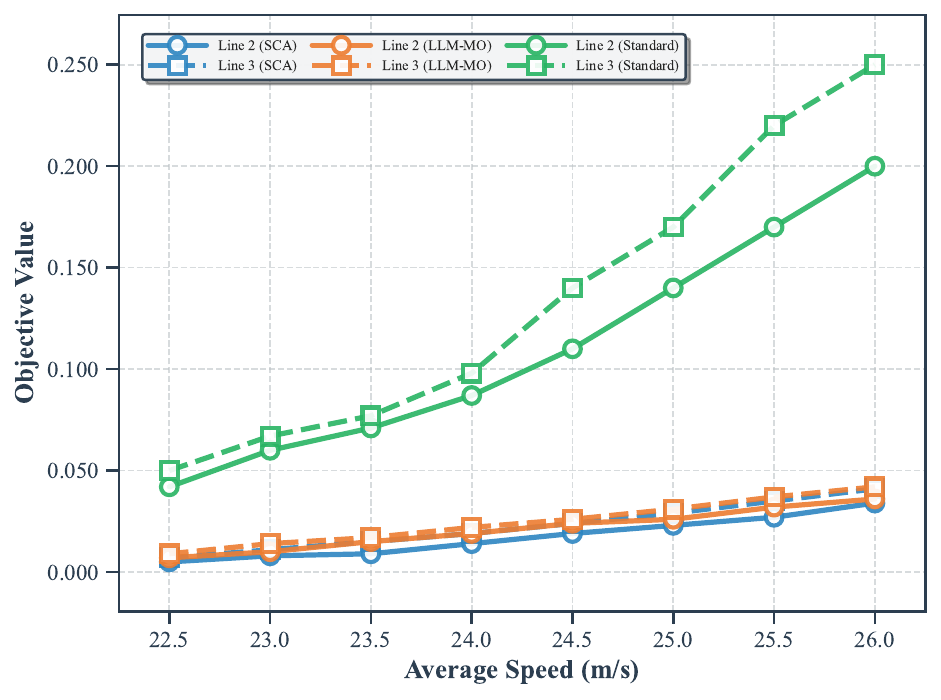}
	\caption{Different Lines' Objective Value Comparison}
	\label{fig5}	
\end{figure}

Fig. \ref{fig4} shows the trend of the selected window size with respect to the average vehicle speed after adopting the two schemes. With a rise in average vehicle speed, we can find that the window size for vehicles on the three lanes decreases, indicating that two schemes adapt the window size based on speed, thus accelerating the vehicle communication speed to reduce AoI. Additionally, we can observe that vehicles with slower speeds have larger window sizes, which reveals that two schemes can modify the window size depending on the vehicle's velocity, allowing faster vehicles to exchange more information and thus achieve fair access.

Fig. \ref{fig5} displays the fairness deviation  for the first $N$ objectives. We can see that vehicles using the two schemes have much smaller fairness differences than those using conventional protocols. This is because the two schemes can effectively tune the window size in order to ensure fair access. Additionally, with a rise in average velocity,  both two schemes' objective values increase. This can be attributed to the fact that faster velocity make fair access more challenging, and it becomes harder to balance the optimization of AoI, leading to a slight increase in fairness deviation. Notably, the second lane have slightly lower fairness deviations, as their speeds tend to be more average, resulting in smaller deviations compared to the overall network.
Meanwhile, the performance of the LLM-MO algorithm is slightly inferior to that of the SCA algorithm, since LLMs struggle to conduct precise modeling and rely mainly on experience-driven reasoning.

\begin{figure}[tb]
	\centering
	\includegraphics[width=\linewidth, scale=1.00]{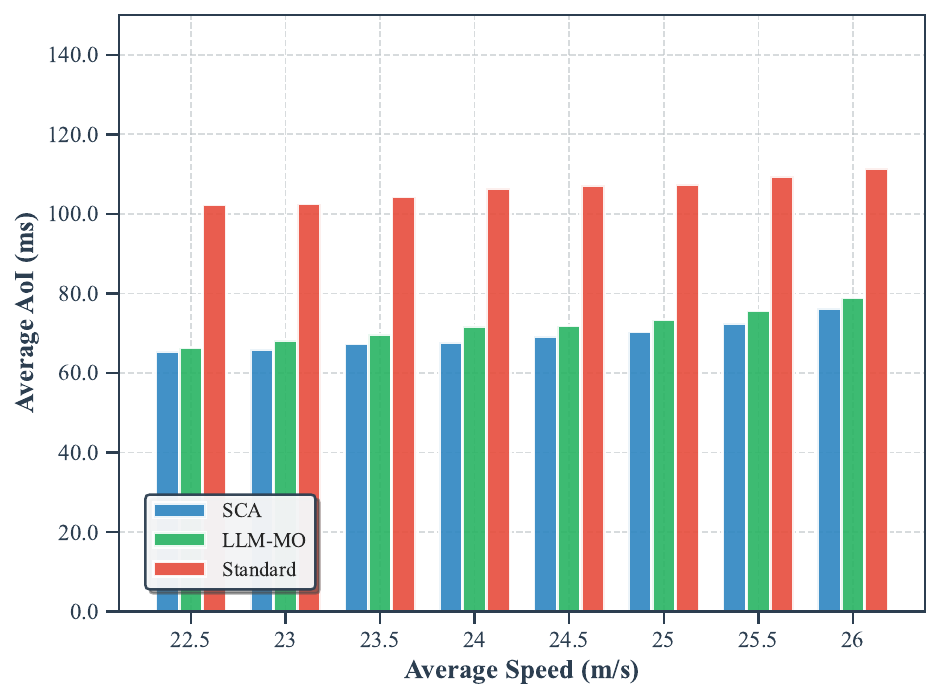}
	\caption{Optimal AoI Comparison}
	\label{fig6}	
\end{figure}
\begin{figure}[tb]
	\centering
	\includegraphics[width=\linewidth, scale=1.00]{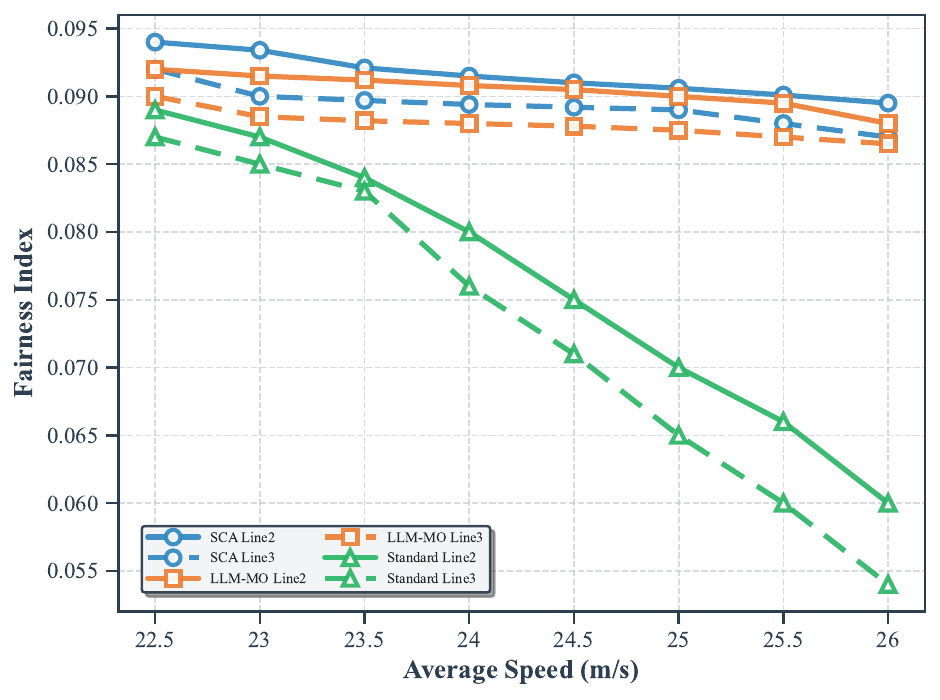}
	\caption{Different Lines' Fairness index Comparison}
	\label{fig7}	
\end{figure}
Fig. \ref{fig6} shows the trend of AoI as the vehicle speed changes. It can be seen that as the speed increases, the AoI for both schemes rises. This is due to the compromises made to achieve both fair access and AoI optimization. However, the AoI under the two schemes are significantly lower than that under conventional protocols, as the two schemes take into account the AoI optimization strategy under re-evaluation mechanism, adjusting  window size according to the speed to minimize AoI. 

The impact of vehicle speed on the fairness index is depicted in Fig. \ref{fig7}. As the average speed increases, the fairness index decreases for both schemes. This is due to the difficulty of optimizing both fair access and AoI when the speed increases. However, we observe that the fairness index for the two schemes remains relatively stable, while the index for the conventional protocol decreases significantly. This indicates that the proposed approach ensures fair access by appropriately tuning the selection window.

\begin{figure}[tb]
	\centering
	\includegraphics[width=\linewidth, scale=1.00]{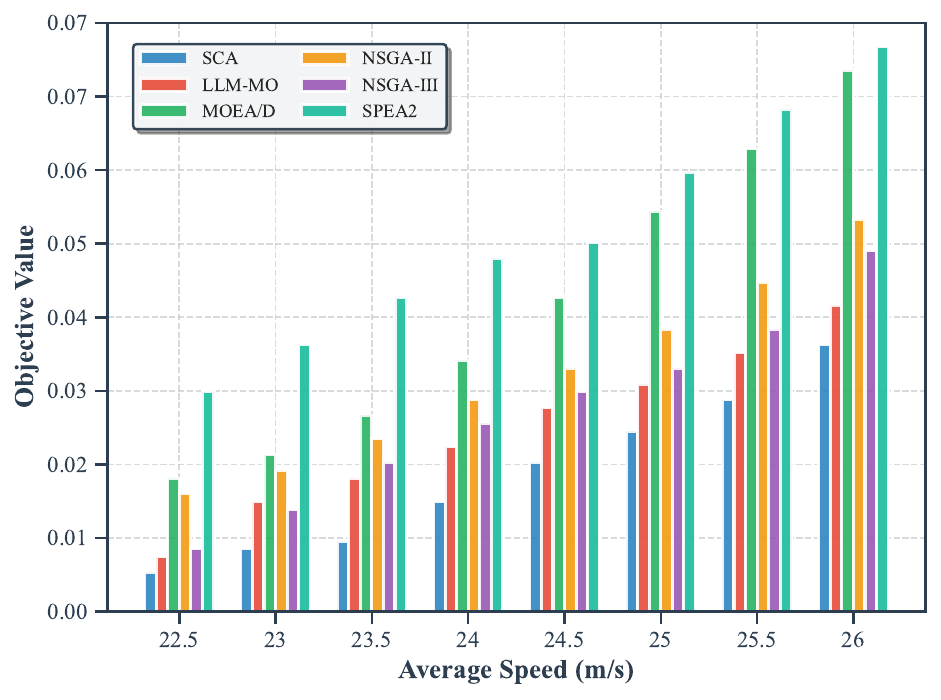}
	\caption{Different algorithms' Objective Value Comparison}
	\label{fig8}	
\end{figure}
\begin{figure}[htbp]
	\centering
	\includegraphics[width=\linewidth, scale=1.00]{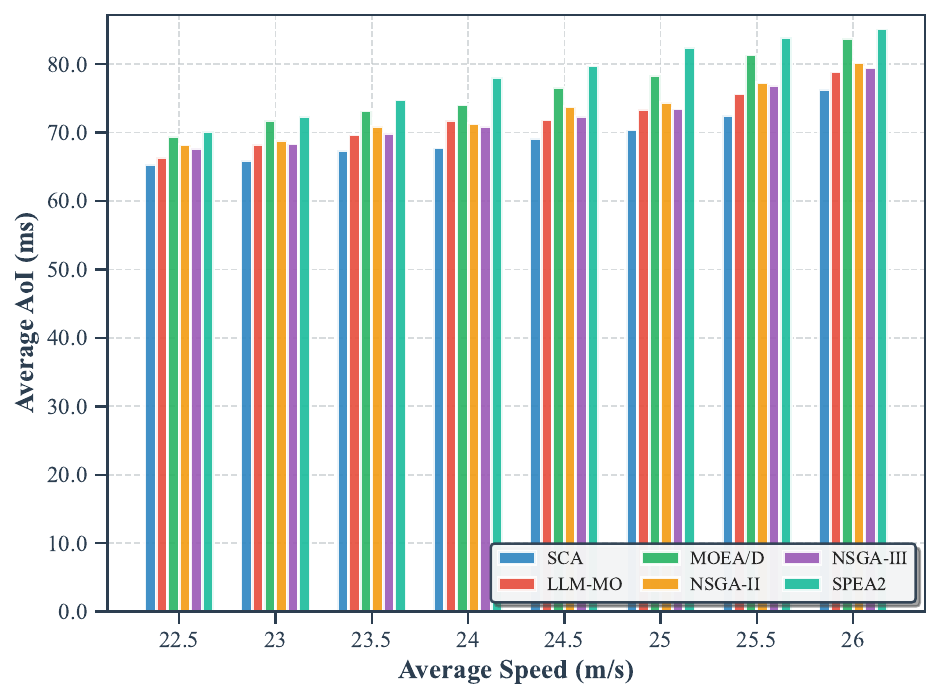}
	\caption{Different algorithms' AoI comparison}
	\label{fig9}	
\end{figure}
\begin{figure}[tb]
	\centering
	\includegraphics[width=\linewidth, scale=1.00]{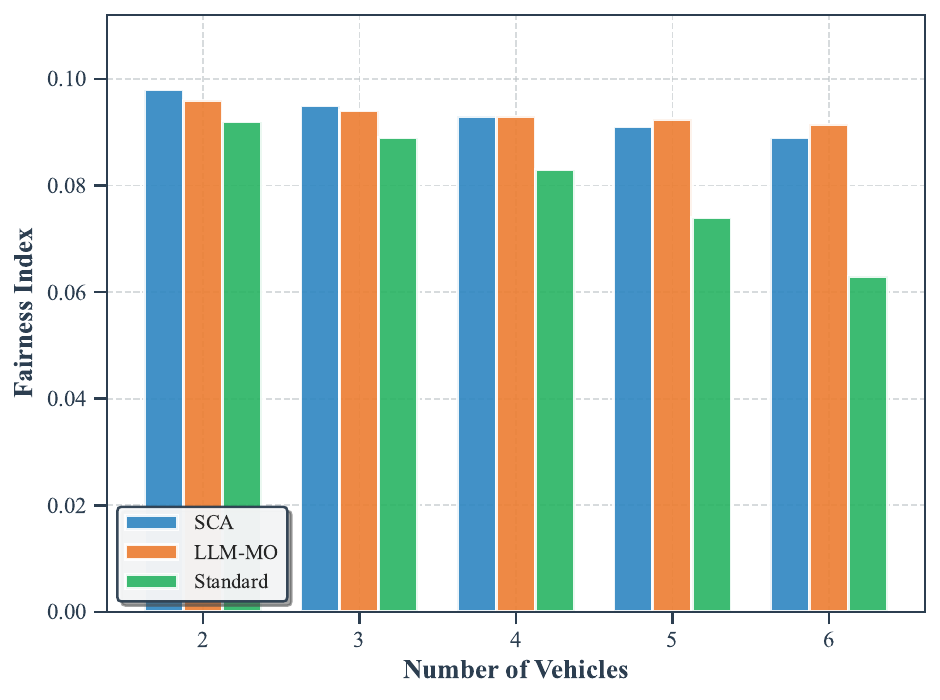}
	\caption{Fairness Index Under Pressure Test }
	\label{fig10}	
\end{figure}
\begin{figure}[tb]
	\centering
	\includegraphics[width=\linewidth, scale=1.00]{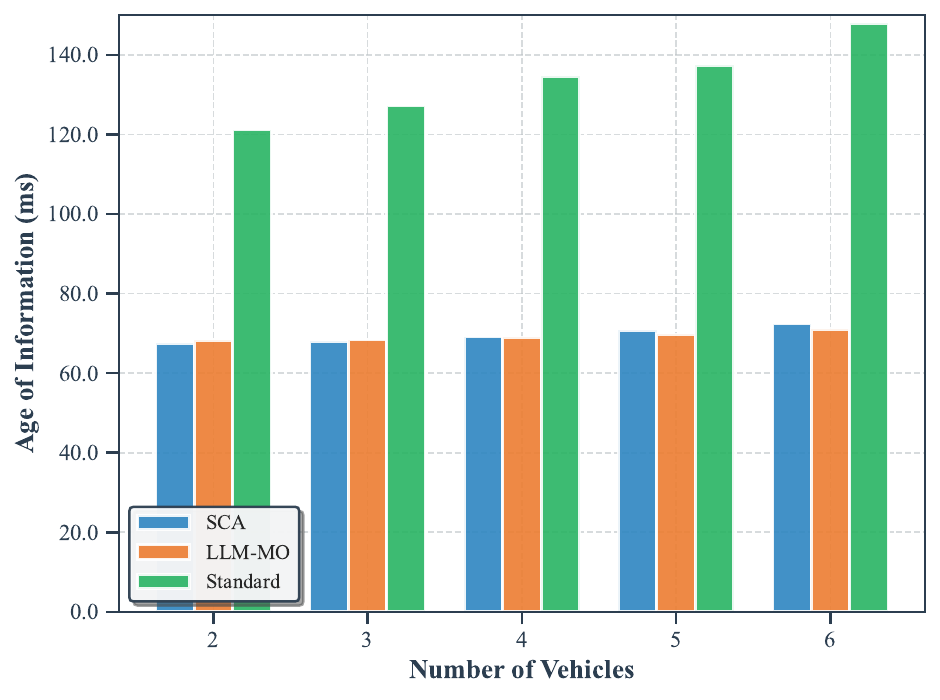}
	\caption{AoI Under Pressure Test}
	\label{fig11}	
\end{figure}
Fig. \ref{fig8} shows the trend of the first $N$ objective values under various benchmark algorithms with a rise in the averaged vehicle velocity. We can see that with vehicle velocity grows, each algorithm demonstrates an upward trend in objective values. However, the increase in the objective value for the two schemes are smaller compared to other algorithms. This is because SCA algorithm uses SCA for convexification and ultimately performs convex optimization to obtain a more accurate global optimum, and the LLM-MO algorithm can obtain offsprings through LLM, while benchmark algorithms require a lot iterations to obtain approximate  solutions.

Fig. \ref{fig9} shows the trend of AoI for different benchmark algorithms with a rise in the averaged vehicle velocity. We can find that with a rise in the averaged vehicle velocity, all the AoI slightly increases due to the difficulty of achieving fair access at higher speeds. However, the two schemes exhibit the lower AoI and the slower growth, as convex optimization can precisely find a solution that balances fair access and AoI and the LLM can obtain better offsprings.

Fig. \ref{fig10} and Fig. \ref{fig11} show the performance of the two schemes under high-pressure. Fig. \ref{fig10} presents the change in the fairness index for different vehicle numbers. We can see that a rise in vehicle number does have a slight impact on the fairness index, as the probability of conflict increases with more vehicles. However, the fairness index under conventional protocols decreases much more sharply. The influence of the number of vehicles on AoI is depicted in Fig. \ref{fig11}. We can see that the increase in vehicle numbers does not significantly affect AoI. The slight increase is due to delays caused by conflicts, indicating that the two schemes exhibit stable performance under high-pressure. Meanwhile, we can find that under high-pressure scenarios, LLM-MO behaves better than SCA. The reason is that under high-dimensional scenarios, the convexification of SCA introduces larger errors, whereas LLM generates offspring through experience-driven reasoning, making it less affected by dimensionality.


\section{Conclusion}
\label{sec8}
In this paper, we consider the problem of fair access. Based on an image semantic communication system, we define a fairness index to measure the extent to which vehicles achieve fair access and implement fair access by adjusting the selection window size. Additionally, we take the new re-evaluation mechanism into consideration. We model AoI using SHS and formulate a multi-objective optimization problem. According to the simulation, we conclude the following findings:

\begin{itemize}
\item [1)] Vehicles with both high and low speeds exhibit large fairness differences, while vehicles with moderate speeds generally show better fair access performance. This is because their fairness index has a smaller difference from the average fairness index due to the moderate speed.
\item [2)] It is difficult to achieve optimal performance for all objectives. As speed increases, achieving fair access optimization may lead to a slight increase in AoI.
\item [3)] SCA algorithm and LLM-MO algorithm are accurate compared to common multi-objective algorithms because they can obtain more precise offsprings.
\end{itemize}

In future work, we might explore the problem of fair access in urban scenarios.


\ifCLASSOPTIONcaptionsoff
  \newpage
\fi

\bibliographystyle{IEEEtran}
\bibliography{IEEEabrv,ref1}

\begin{thebibliography}{10}
\providecommand{\url}[1]{#1}
\csname url@samestyle\endcsname
\providecommand{\newblock}{\relax}
\providecommand{\bibinfo}[2]{#2}
\providecommand{\BIBentrySTDinterwordspacing}{\spaceskip=0pt\relax}
\providecommand{\BIBentryALTinterwordstretchfactor}{4}
\providecommand{\BIBentryALTinterwordspacing}{\spaceskip=\fontdimen2\font plus
\BIBentryALTinterwordstretchfactor\fontdimen3\font minus
  \fontdimen4\font\relax}
\providecommand{\BIBforeignlanguage}[2]{{%
\expandafter\ifx\csname l@#1\endcsname\relax
\typeout{** WARNING: IEEEtran.bst: No hyphenation pattern has been}%
\typeout{** loaded for the language `#1'. Using the pattern for}%
\typeout{** the default language instead.}%
\else
\language=\csname l@#1\endcsname
\fi
#2}}
\providecommand{\BIBdecl}{\relax}
\BIBdecl

\bibitem{intro1}
X.~Zhang, J.~Li, J.~Zhou, S.~Zhang, J.~Wang, Y.~Yuan, J.~Liu, and J.~Li,
  ``Vehicle-to-everything communication in intelligent connected vehicles: A
  survey and taxonomy,'' \emph{Automotive Innovation}, pp. 1--33, 2025.

\bibitem{arx1}
Q.~Wu and J.~Zheng, ``Performance modeling of ieee 802.11 dcf based fair
  channel access for vehicular-to-roadside communication in a non-saturated
  state,'' in \emph{2014 IEEE International Conference on Communications
  (ICC)}, 2014, pp. 2575--2580.

\bibitem{arx2}
Z.~Shao, Q.~Wu, P.~Fan, N.~Cheng, Q.~Fan, and J.~Wang, ``Semantic-aware
  resource allocation based on deep reinforcement learning for 5g-v2x
  hetnets,'' \emph{IEEE Communications Letters}, vol.~28, no.~10, pp.
  2452--2456, 2024.

\bibitem{intro2}
S.~Feng, X.~Yan, H.~Sun, Y.~Feng, and H.~X. Liu, ``Intelligent driving
  intelligence test for autonomous vehicles with naturalistic and adversarial
  environment,'' \emph{Nature communications}, vol.~12, no.~1, p. 748, 2021.

\bibitem{arx3}
K.~Qi, Q.~Wu, P.~Fan, N.~Cheng, W.~Chen, and K.~B. Letaief,
  ``Reconfigurable-intelligent-surface-aided vehicular edge computing: Joint
  phase-shift optimization and multiuser power allocation,'' \emph{IEEE
  Internet of Things Journal}, vol.~12, no.~1, pp. 764--777, 2025.

\bibitem{arx4}
\BIBentryALTinterwordspacing
S.~Song, Z.~Zhang, Q.~Wu, P.~Fan, and Q.~Fan, ``Joint optimization of age of
  information and energy consumption in nr-v2x system based on deep
  reinforcement learning,'' \emph{Sensors}, vol.~24, no.~13, 2024. [Online].
  Available: \url{https://www.mdpi.com/1424-8220/24/13/4338}
\BIBentrySTDinterwordspacing

\bibitem{intro3}
C.~Ma, J.~Zhu, M.~Liu, H.~Zhao, N.~Liu, and X.~Zou, ``Parking edge computing:
  Parked-vehicle-assisted task offloading for urban vanets,'' \emph{IEEE
  Internet of Things Journal}, vol.~8, no.~11, pp. 9344--9358, 2021.

\bibitem{intro4}
H.~Yang, Z.~Wei, Z.~Feng, X.~Chen, Y.~Li, and P.~Zhang, ``Intelligent
  computation offloading for mec-based cooperative vehicle infrastructure
  system: A deep reinforcement learning approach,'' \emph{IEEE Transactions on
  Vehicular Technology}, vol.~71, no.~7, pp. 7665--7679, 2022.

\bibitem{intro5}
I.~Soto, M.~Calderon, O.~Amador, and M.~Urue{\~n}a, ``A survey on road safety
  and traffic efficiency vehicular applications based on c-v2x technologies,''
  \emph{Vehicular Communications}, vol.~33, p. 100428, 2022.

\bibitem{aa1}
X.~Wang, K.~Tao, N.~Cheng, Z.~Yin, Z.~Li, Y.~Zhang, and X.~Shen, ``Radiodiff:
  An effective generative diffusion model for sampling-free dynamic radio map
  construction,'' \emph{IEEE Transactions on Cognitive Communications and
  Networking}, vol.~11, no.~2, pp. 738--750, 2025.

\bibitem{intro6}
J.~Wang, W.~Chai, A.~Venkatachalapathy, K.~L. Tan, A.~Haghighat,
  S.~Velipasalar, Y.~Adu-Gyamfi, and A.~Sharma, ``A survey on driver behavior
  analysis from in-vehicle cameras,'' \emph{IEEE Transactions on Intelligent
  Transportation Systems}, vol.~23, no.~8, pp. 10\,186--10\,209, 2021.

\bibitem{arx11}
P.~Fan, C.~Feng, Y.~Wang, and N.~Ge, ``Investigation of the time-offset-based
  qos support with optical burst switching in wdm networks,'' in \emph{2002
  IEEE International Conference on Communications. Conference Proceedings. ICC
  2002 (Cat. No.02CH37333)}, vol.~5, 2002, pp. 2682--2686 vol.5.

\bibitem{intro7}
X.~Luo, H.-H. Chen, and Q.~Guo, ``Semantic communications: Overview, open
  issues, and future research directions,'' \emph{IEEE Wireless
  communications}, vol.~29, no.~1, pp. 210--219, 2022.

\bibitem{intro8}
Y.~Shao, Q.~Cao, and D.~G{\"u}nd{\"u}z, ``A theory of semantic communication,''
  \emph{IEEE Transactions on Mobile Computing}, vol.~23, no.~12, pp.
  12\,211--12\,228, 2024.

\bibitem{arx5}
X.~Wang, Q.~Wu, P.~Fan, Q.~Fan, H.~Zhu, and J.~Wang, ``Vehicle selection for
  c-v2x mode 4-based federated edge learning systems,'' \emph{IEEE Systems
  Journal}, vol.~18, no.~4, pp. 1927--1938, 2024.

\bibitem{5G}
M.~H.~C. Garcia, A.~Molina-Galan, M.~Boban, J.~Gozalvez, B.~Coll-Perales,
  T.~{\c{S}}ahin, and A.~Kousaridas, ``A tutorial on {5G NR V2X}
  communications,'' \emph{IEEE Communications Surveys \& Tutorials}, vol.~23,
  no.~3, pp. 1972--2026, 2021.

\bibitem{intro9}
M.~M. Saad, M.~T.~R. Khan, S.~H.~A. Shah, and D.~Kim, ``Advancements in
  vehicular communication technologies: C-v2x and nr-v2x comparison,''
  \emph{IEEE Communications Magazine}, vol.~59, no.~8, pp. 107--113, 2021.

\bibitem{arx6}
C.~Meng, K.~Xiong, W.~Chen, B.~Gao, P.~Fan, and K.~B. Letaief, ``Sum-rate
  maximization in star-ris-assisted rsma networks: A ppo-based algorithm,''
  \emph{IEEE Internet of Things Journal}, vol.~11, no.~4, pp. 5667--5680, 2024.

\bibitem{intro10}
A.~Molina-Galan, L.~Lusvarghi, B.~Coll-Perales, J.~Gozalvez, and M.~L. Merani,
  ``On the impact of re-evaluation in 5g nr v2x mode 2,'' \emph{IEEE
  Transactions on Vehicular Technology}, vol.~73, no.~2, pp. 2669--2683, 2023.

\bibitem{arx7}
W.~Mao, K.~Xiong, Y.~Lu, P.~Fan, and Z.~Ding, ``Energy consumption minimization
  in secure multi-antenna uav-assisted mec networks with channel uncertainty,''
  \emph{IEEE Transactions on Wireless Communications}, vol.~22, no.~11, pp.
  7185--7200, 2023.

\bibitem{arx8}
Y.~Dong, Z.~Chen, S.~Liu, P.~Fan, and K.~B. Letaief, ``Age-upon-decisions
  minimizing scheduling in internet of things: To be random or to be
  deterministic?'' \emph{IEEE Internet of Things Journal}, vol.~7, no.~2, pp.
  1081--1097, 2020.

\bibitem{intro11}
M.~M. Saad, M.~A. Tariq, J.~Seo, M.~Ajmal, and D.~Kim, ``Age-of-information
  aware intelligent mac for congestion control in nr-v2x,'' in \emph{2023
  Fourteenth International Conference on Ubiquitous and Future Networks
  (ICUFN)}.\hskip 1em plus 0.5em minus 0.4em\relax IEEE, 2023, pp. 265--270.

\bibitem{arx9}
X.~Di, K.~Xiong, P.~Fan, H.-C. Yang, and K.~B. Letaief, ``Optimal resource
  allocation in wireless powered communication networks with user
  cooperation,'' \emph{IEEE Transactions on Wireless Communications}, vol.~16,
  no.~12, pp. 7936--7949, 2017.

\bibitem{arx10}
T.~Li, P.~Fan, Z.~Chen, and K.~B. Letaief, ``Optimum transmission policies for
  energy harvesting sensor networks powered by a mobile control center,''
  \emph{IEEE Transactions on Wireless Communications}, vol.~15, no.~9, pp.
  6132--6145, 2016.

\bibitem{fair1}
T.~T. Nguyen, K.~Elbassioni, N.~C. Luong, D.~Niyato, and D.~I. Kim, ``Access
  management in joint sensing and communication systems: Efficiency versus
  fairness,'' \emph{IEEE Transactions on Vehicular Technology}, vol.~71, no.~5,
  pp. 5128--5142, 2022.

\bibitem{fair2}
\BIBentryALTinterwordspacing
F.~Mehmeti and W.~Kellerer, ``Max-min fair resource allocation in sd-ran,'' in
  \emph{Proceedings of the 18th ACM International Symposium on QoS and Security
  for Wireless and Mobile Networks}, ser. Q2SWinet '22.\hskip 1em plus 0.5em
  minus 0.4em\relax New York, NY, USA: Association for Computing Machinery,
  2022, p. 27–35. [Online]. Available:
  \url{https://doi.org/10.1145/3551661.3561359}
\BIBentrySTDinterwordspacing

\bibitem{fair3}
T.~Wang and R.~Adve, ``Fair licensed spectrum sharing between two mnos using
  resource optimization in multi-cell multi-user mimo networks,'' \emph{IEEE
  Transactions on Wireless Communications}, vol.~21, no.~8, pp. 6714--6730,
  2022.

\bibitem{wan}
Q.~Wu, Z.~Wan, Q.~Fan, P.~Fan, and J.~Wang, ``Velocity-adaptive access scheme
  for mec-assisted platooning networks: Access fairness via data freshness,''
  \emph{IEEE Internet of Things Journal}, vol.~9, no.~6, pp. 4229--4244, 2021.

\bibitem{fair4}
W.~Qiong, S.~Shuai, W.~Ziyang, F.~Qiang, F.~Pingyi, and Z.~Cui, ``Towards v2i
  age-aware fairness access: A dqn based intelligent vehicular node training
  and test method,'' \emph{Chinese Journal of Electronics}, vol.~32, no.~6, pp.
  1230--1244, 2023.

\bibitem{reaoi1}
A.~Rolich, M.~Yildiz, I.~Turcanu, A.~Vinel, and A.~Baiocchi, ``On the trade-off
  between aoi performance and resource reuse efficiency in 5g nr v2x
  sidelink,'' in \emph{2025 IEEE Vehicular Networking Conference (VNC)}.\hskip
  1em plus 0.5em minus 0.4em\relax IEEE, 2025, pp. 1--8.

\bibitem{reaoi2}
A.~Rolich, I.~Turcanu, and A.~Baiocchi, ``Aoi-aware and persistence-driven
  congestion control in 5g nr-v2x sidelink communications,'' in \emph{2024 22nd
  Mediterranean Communication and Computer Networking Conference
  (MedComNet)}.\hskip 1em plus 0.5em minus 0.4em\relax IEEE, 2024, pp. 1--4.

\bibitem{reaoi3}
E.~Markova, V.~Manaeva, E.~Zhbankova, D.~Moltchanov, P.~Balabanov,
  Y.~Koucheryavy, and Y.~Gaidamaka, ``Performance-utilization trade-offs for
  state update services in 5g nr systems,'' \emph{IEEE Access}, 2024.

\bibitem{reaoi4}
X.~Wang, K.~Chen, K.~Li, and Q.~Yang, ``Aoi-aware data propagation in edge
  computing assisted nr-v2x networks,'' in \emph{Proceedings of the 6th
  Asia-Pacific Workshop on Networking}, 2022, pp. 103--104.

\bibitem{reaoi5}
S.~Song, Z.~Zhang, Q.~Wu, P.~Fan, and Q.~Fan, ``Joint optimization of age of
  information and energy consumption in nr-v2x system based on deep
  reinforcement learning,'' \emph{Sensors}, vol.~24, no.~13, p. 4338, 2024.

\bibitem{semantic}
W.~Zhang, Y.~Wang, M.~Chen, T.~Luo, and D.~Niyato, ``Optimization of image
  transmission in cooperative semantic communication networks,'' \emph{IEEE
  Transactions on Wireless Communications}, vol.~23, no.~2, pp. 861--873, 2023.

\bibitem{semantic2}
K.~Tang, Y.~Niu, J.~Huang, J.~Shi, and H.~Zhang, ``Unbiased scene graph
  generation from biased training,'' in \emph{2020 IEEE/CVF Conference on
  Computer Vision and Pattern Recognition (CVPR)}, 2020, pp. 3713--3722.

\bibitem{armodel}
\BIBentryALTinterwordspacing
S.~Kirthiga and M.~Jayakumar, ``Autoregressive channel modeling and estimation
  using kalman filter for downlink lte systems,'' in \emph{Proceedings of the
  1st Amrita ACM-W Celebration on Women in Computing in India}, ser. A2CWiC
  '10.\hskip 1em plus 0.5em minus 0.4em\relax New York, NY, USA: Association
  for Computing Machinery, 2010. [Online]. Available:
  \url{https://doi.org/10.1145/1858378.1858423}
\BIBentrySTDinterwordspacing

\bibitem{prr1}
C.~Brady, L.~Cao, and S.~Roy, ``Modeling of nr c-v2x mode 2 throughput,'' in
  \emph{2022 IEEE International Workshop Technical Committee on Communications
  Quality and Reliability (CQR)}.\hskip 1em plus 0.5em minus 0.4em\relax IEEE,
  2022, pp. 19--24.

\bibitem{dataset}
\BIBentryALTinterwordspacing
R.~Krishna, Y.~Zhu, O.~Groth, J.~Johnson, K.~Hata, J.~Kravitz, S.~Chen,
  Y.~Kalantidis, L.-J. Li, D.~A. Shamma, M.~S. Bernstein, and F.-F. Li,
  ``Visual genome: Connecting language and vision using crowdsourced dense
  image annotations,'' 2016. [Online]. Available:
  \url{https://arxiv.org/abs/1602.07332}
\BIBentrySTDinterwordspacing

\bibitem{aoi1}
A.~Maatouk, M.~Assaad, and A.~Ephremides, ``On the age of information in a csma
  environment,'' \emph{IEEE/ACM Transactions on Networking}, vol.~28, no.~2,
  pp. 818--831, 2020.

\bibitem{aoi2}
R.~D. Yates and S.~K. Kaul, ``The age of information: Real-time status updating
  by multiple sources,'' \emph{IEEE Transactions on Information Theory},
  vol.~65, no.~3, pp. 1807--1827, 2019.

\bibitem{latency}
M.~C. Lucas-Esta{\~n}, B.~Coll-Perales, T.~Shimizu, J.~Gozalvez, T.~Higuchi,
  S.~Avedisov, O.~Altintas, and M.~Sepulcre, ``An analytical latency model and
  evaluation of the capacity of {5G NR} to support {V2X} services using {V2N2V}
  communications,'' \emph{IEEE Transactions on Vehicular Technology}, vol.~72,
  no.~2, pp. 2293--2306, 2022.

\bibitem{SCA}
A.~Beck, A.~Ben-Tal, and L.~Tetruashvili, ``A sequential parametric convex
  approximation method with applications to nonconvex truss topology design
  problems,'' \emph{Journal of Global Optimization}, vol.~47, no.~1, pp.
  29--51, 2010.

\bibitem{LLM}
F.~Liu, X.~Lin, S.~Yao, Z.~Wang, X.~Tong, M.~Yuan, and Q.~Zhang, ``Large
  language model for multiobjective evolutionary optimization,'' in
  \emph{International Conference on Evolutionary Multi-Criterion
  Optimization}.\hskip 1em plus 0.5em minus 0.4em\relax Springer, 2025, pp.
  178--191.

\bibitem{3gpp}
\BIBentryALTinterwordspacing
3GPP, ``{Release 16 Description; Summary of Rel-16 Work Items},'' {3rd
  Generation Partnership Project (3GPP)}, Technical report (TR) 21.916, 04
  2020, version 16.2.0. [Online]. Available:
  \url{https://portal.3gpp.org/desktopmodules/Specifications/SpecificationDetails.aspx?specificationId=3493}
\BIBentrySTDinterwordspacing

\bibitem{ADMM}
\BIBentryALTinterwordspacing
S.~Boyd, N.~Parikh, E.~Chu, B.~Peleato, and J.~Eckstein, ``Distributed
  optimization and statistical learning via the alternating direction method of
  multipliers,'' \emph{Foundations and Trends® in Machine Learning}, vol.~3,
  no.~1, pp. 1--122, 2011. [Online]. Available:
  \url{http://dx.doi.org/10.1561/2200000016}
\BIBentrySTDinterwordspacing

\bibitem{weight}
R.~T. Marler and J.~S. Arora, ``The weighted sum method for multi-objective
  optimization: new insights,'' \emph{Structural and multidisciplinary
  optimization}, vol.~41, no.~6, pp. 853--862, 2010.

\bibitem{moea1}
V.~L. Vachhani, V.~K. Dabhi, and H.~B. Prajapati, ``Survey of multi objective
  evolutionary algorithms,'' in \emph{2015 International Conference on
  Circuits, Power and Computing Technologies [ICCPCT-2015]}, 2015, pp. 1--9.

\bibitem{pymoo}
J.~{Blank} and K.~{Deb}, ``pymoo: Multi-objective optimization in python,''
  \emph{IEEE Access}, vol.~8, pp. 89\,497--89\,509, 2020.

\bibitem{cvxpy}
\BIBentryALTinterwordspacing
S.~Diamond and S.~Boyd, ``{CVXPY}: A {P}ython-embedded modeling language for
  convex optimization,'' \emph{Journal of Machine Learning Research}, 2016, to
  appear. [Online]. Available:
  \url{https://stanford.edu/~boyd/papers/pdf/cvxpy_paper.pdf}
\BIBentrySTDinterwordspacing

\end{thebibliography}


\end{document}